\documentclass[acmtog, authorversion, nonacm]{acmart}

\AtBeginDocument{%
  \providecommand\BibTeX{{%
    \normalfont B\kern-0.5em{\scshape i\kern-0.25em b}\kern-0.8em\TeX}}}

\acmJournal{FACMP}

\usepackage{xfp}
\usepackage{wrapfig}
\usepackage[percent]{overpic}
\usepackage{mathtools}
\usepackage{booktabs}
\usepackage{amsmath}
\usepackage{relsize}
\usepackage{xcolor}
\usepackage{caption}
\usepackage{dsfont}
\usepackage{enumitem}
\usepackage{subcaption}
\usepackage{tikz}
\usetikzlibrary{tikzmark,calc}
\usepackage{pifont}
\usepackage{stackengine}
\usepackage{bbding}

\newif\ifdraft
\draftfalse

\ifdraft
\newcommand{\dcc}[1]{{\color{red}[\textbf{DC:} #1]}}
\newcommand{\ahc}[1]{{\color{teal}[\textbf{AH:} #1]}}
\newcommand{\rgc}[1]{{\color{orange}[\textbf{RG:} #1]}}
\newcommand{\opc}[1]{{\color{blue}[\textbf{OP} #1]}}

\newcommand{\ah}[1]{{\color{teal}#1}}

\else
\newcommand{\dcc}[1]{}
\newcommand{\rgc}[1]{}
\newcommand{\ahc}[1]{}
\newcommand{\opc}[1]{}
\newcommand{\osc}[1]{}

\newcommand{\ah}[1]{{\color{black}#1}}

\fi

\newcommand{\ourmethod}{SPAGHETTI}
\newcommand{\set}[3]{\{#1_{#2}\}_{#2=1}^{#3}}
\newcommand{\reals}{\mathds{R}}

\definecolor{mygray}{RGB}{140, 140, 140}
\newcommand{\gray}[1]{{\color{mygray}#1}}

\definecolor{dashedpurple}{RGB}{118,0,103}
\definecolor{dashedorange}{RGB}{181,100,13}

\newcommand{\za}{\mathbf{z}^{a}}
\newcommand{\zA}{Z^{a}}
\newcommand{\zb}{\mathbf{z}^{b}}
\newcommand{\zbh}{\hat{\mathbf{z}}^{b}}
\newcommand{\zB}{Z^{b}}
\newcommand{\zBh}{\hat{Z}^{b}}
\newcommand{\zc}{\mathbf{z}^{c}}
\newcommand{\zC}{Z^{c}}

\newcommand{\gj}{\mathbf{g}_{j}}
\newcommand{\gjhat}{\hat{\mathbf{g}_{j}}}
\newcommand{\sj}{\mathbf{s}_{j}}

\newcommand{\netA}{f_a}
\newcommand{\netB}{f_b}
\newcommand{\netC}{f_c}
\newcommand{\dm}{d_\textrm{model}}
\newcommand{\ds}{d_\textrm{surf}}
\newcommand{\dpe}{d_\textrm{pe}}

\newcommand{\xcoord}{\mathbf{x}}
\newcommand{\xcoordhat}{\hat{\mathbf{x}}}
\newcommand{\xin}{X_{\textrm{vol}}}
\newcommand{\xbatch}{X_{\textrm{surf}}}

\DeclareMathOperator*{\argmax}{arg\,max}
\DeclareMathOperator*{\argmin}{arg\,min}

\newcommand{\ts}{@{\hskip 1\tabcolsep}}
\newcommand{\tss}{@{\hskip .4\tabcolsep}}
\newcommand{\tsm}{@{\hskip 2\tabcolsep}}
\newcommand{\tsl}{@{\hskip 2.5\tabcolsep}}

\newcommand{\noenc}{\ourmethod{} no-enc}
\newcommand{\nodis}{\ourmethod{} no-dis}

\newcommand{\cdemd}{\scriptsize{ CD / EMD}}
\newcommand{\downa}{\scriptsize{$\downarrow$}}
\newcommand{\upa}{\scriptsize{$\uparrow$}}

\newcommand{\iouall}{Cov. }
\newcommand{\ioupart}{Cor.}

\setcitestyle{nosort,square}
\begin{document}

\title{SPAGHETTI: Editing Implicit Shapes Through Part Aware Generation}

\author{Amir Hertz}
\affiliation{\institution{Tel Aviv University} }
\author{Or Perel}
\affiliation{\institution{Tel Aviv University} } \affiliation{\institution{NVIDIA}}
\author{Raja Giryes}
\affiliation{\institution{Tel Aviv University}} 
\author{Olga Sorkine-Hornung}
\affiliation{\institution{ETH Zurich, Switzerland}}
\author{Daniel Cohen-Or}
\affiliation{\institution{Tel Aviv University} }

\begin{abstract}
Neural implicit fields are quickly emerging as an attractive representation for learning based techniques.
However, adopting them for 3D shape modeling and editing is challenging. We introduce a method for \textbf{E}diting \textbf{I}mplicit \textbf{S}hapes \textbf{T}hrough \textbf{P}art \textbf{A}ware \textbf{G}enera\textbf{T}ion, permuted in short as SPAGHETTI. Our architecture allows for manipulation of implicit shapes by means of transforming, interpolating and combining shape segments together, without requiring explicit part supervision. SPAGHETTI disentangles shape part representation into extrinsic and intrinsic geometric information. This characteristic enables a generative framework with part-level control. The modeling capabilities of SPAGHETTI are demonstrated using an interactive graphical interface, where users can directly edit neural implicit shapes.
Our code, pre-trained models and editing user interface are available at 
\textit{\color{magenta}\url{https://github.com/amirhertz/spaghetti}}.
\end{abstract}

\begin{teaserfigure}
  \includegraphics[width=\textwidth]{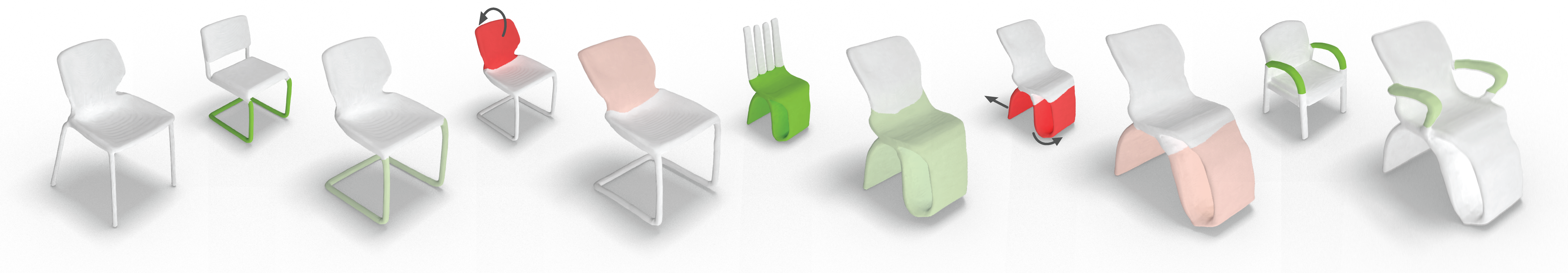}
  \caption{Design exploration with \ourmethod{}. Using our method, the user can easily compose new shapes (in grey) out of parts (in green) taken from reference shapes or make further local adjustments of selected parts (in red). }
  \label{fig:teaser}
\end{teaserfigure}

\maketitle

\newcommand\offsetx{.93}
\newcommand\offsety{.39}

\begin{tikzpicture}[remember picture, overlay]
\node (init) {};
\draw[line width=0.15mm] (\fpeval{\offsetx - .95},\fpeval{\offsety + 13.15}) .. controls +(.6,.4) and +(-.6,.4) .. (\fpeval{\offsetx + 1.1},\fpeval{\offsety +13.15});
\draw[line width=0.25mm] (\fpeval{\offsetx -.55},\fpeval{\offsety + 13.15}) .. controls +(.4,.25) and +(-.4,.25) .. (\fpeval{\offsetx + .7},\fpeval{\offsety + 13.15});
\draw[line width=0.25mm] (\fpeval{\offsetx + .3}, \fpeval{\offsety + 13.15}) .. controls +(.4,.25) and +(-.4,.25) .. (\fpeval{\offsetx + 1.55},\fpeval{\offsety + 13.15});
\draw[line width=0.3mm] (\fpeval{\offsetx -.62},\fpeval{\offsety +12.65}) .. controls +(.4,-.2) and +(-.4,-.2) .. (\fpeval{\offsetx + 1.4},\fpeval{\offsety +12.65});
\draw[line width=0.15mm] (\fpeval{\offsetx -.89},\fpeval{\offsety +12.65}) .. controls +(.5,-.35) and +(-.5,-.35) .. (\fpeval{\offsetx + 2},\fpeval{\offsety + 12.65});
\end{tikzpicture}
\vspace{-5mm}

\section{Introduction}

In recent years, there is a surge of interest in applying neural implicit fields to represent 3D shapes and scenes. By design, such parameterizations do not limit the shape resolution, thereby allowing to faithfully recover the underlying continuous surface or volume.

The learned nature of neural implicit fields also promotes them as naturally compressed representations, capable of capturing high resolution details with low memory cost. Altogether, these attributes make them an intriguing medium for developing novel generative techniques.

Most of the recent research efforts have been focused on refining the quality of represented signals \cite{sitzmann2019siren, takikawa2021neural, martel2021acorn} or leveraging implicit representations for shape reconstruction \cite{erler2020points2surf,genova2020local, chabra2020deep}. While implicit surface representations are well established in classical shape modeling literature (e.g., \cite{Cani:ShapeModelingBook:2008,schmidt2011shapeshop}), so far only little attention has been given to editing of neural implicit shapes \cite{hao2020dualsdf}. 
In particular, conditioning the form of a neural implicit shape on specified user controls is not straightforward, further hindering the adoption in 3D creative applications.

In this paper, we introduce \ourmethod{}, a novel generative model that supports direct editing of neural implicit shapes. Our framework allows for part level of control by (i) applying transformations on local areas of the generated object; and (ii) mixing and interpolating segments of different shapes.

The editing power of our model comes from its dual-level disentanglement. First, our network learns to separate local part representations from each other. This is essential, as modifications to a single part should have little effect on the rest. Our network is trained to achieve this separation without explicit part supervision. Second, each part representation is factored into intrinsic and extrinsic components, respectively controlling its detailed surface geometry and embedding in 3D space. Doing so allows our learning process to introduce local affine transformations on shape parts while keeping them within the data distribution (see Figure~\ref{fig:teaser}).

\begin{figure}
\centering

    \begin{overpic}[width=1\columnwidth,tics=10, trim=0mm 0 0mm 0,clip]{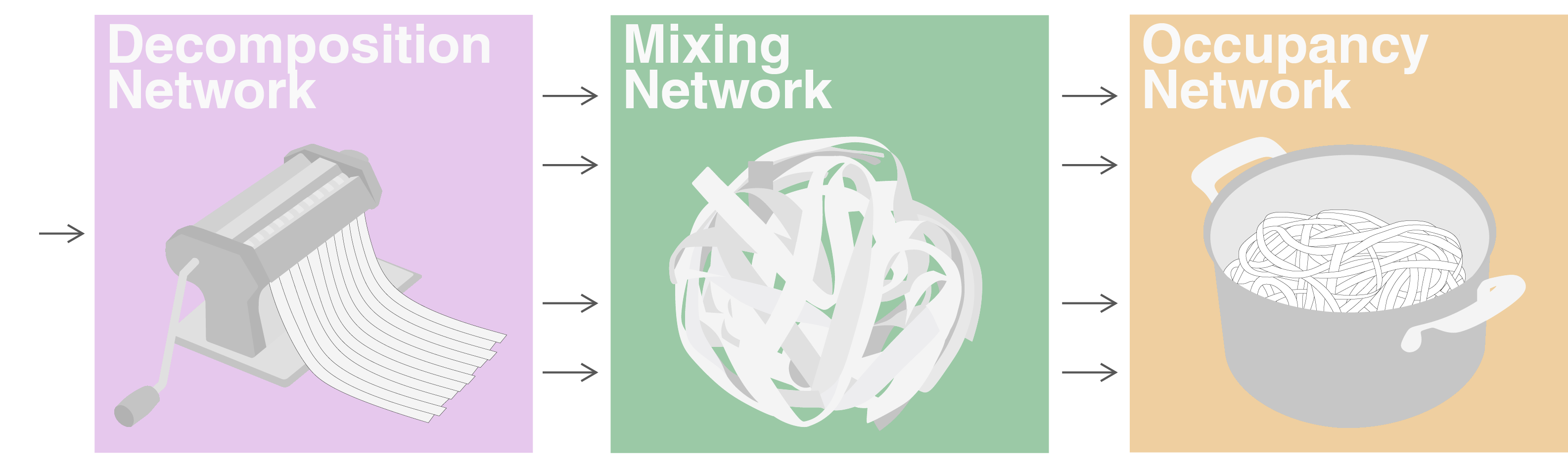}
    
     \put(-.8,14){$\mathbf{\za}$}
      
    \put(34.6,13.4){$\mathbf{\zB}$}
 
     \put(67.9,13.4){$\mathbf{\zC}$}
     
     \put(18.1,-2.2){$\netA$}
     \put(52,-2.2){$\netB$}
     \put(84.5,-2.2){$\netC$}

    \end{overpic}
    
\caption{Method overview. Our implicit shape generative model learns to decompose a shape embedding ${\za}$ into distinct embeddings $\zB$ that correspond to distinct 3D parts. Then, a mixing network outputs contextual embeddings $\zC$. Finally the implicit shape is given by a third occupancy network $\netC$ that is conditioned on the contextual part embeddings.}

\label{fig:overview}

\end{figure}
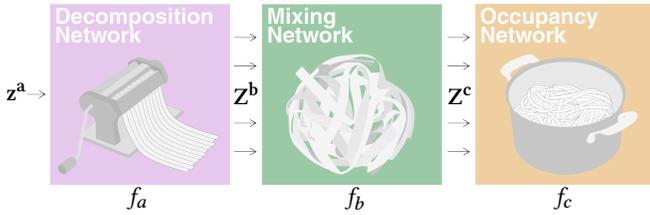

As illustrated in Figure~\ref{fig:overview}, our architecture can be roughly divided to three steps. At the beginning of our pipeline, the Decomposition network receives a latent shape embedding ${\za}$ and projects it onto a set of latent codes $\zB$. Each $\zb \in \zB$ corresponds a distinct part of the 3D shape, with its latent representation of surface information and global transformation. 
Then, the Mixing network, based on a transformer encoder architecture, processes $\zB$ and outputs contextual codes $\zC$. Finally, the implicit shape is generated
by an Occupancy network, in the form of a transformer decoder, where a query coordinate is weighted according to $\zC$ to output its occupancy value.

During inference, the user can control the 3D shape by modifying its components in their raw latent state. The user can modify the local extrinsic properties of each part, as well as  add, remove or replace components taken from other shapes. After each editing step, the modified parts are re-composed into a new implicit shape. See examples in Figure~\ref{fig:teaser}.

To enable the editing of new shapes that were not seen during training, we introduce a \textit{shape inversion} optimization, which finds the matching part codes for a given unseen shape. Our mid latent representation of disentangled part embeddings facilitates the extrapolation outside the training data and enables high quality inversions.

We demonstrate the effectiveness of our method using a graphical user interface, where \ourmethod{} is accelerated by an Octree, to allow for an interactive editing experience of 3D implicit shapes.

\section{Related Work}

\textit{3D generative models.} 
The introduction of 
deep generative models within the computer vision community \cite{Kingma2014vae, goodfellow2014gans, dinh2014nice, radford2016dcgan, vandenoord2016pixelcnn} quickly spawned a rich line of works capable of producing visually appealing images. Despite their success on images, adapting these models to generate 3D shapes proved to be non-trivial. \citet{wu20163dgan} were the first to extend the unconditional generative adversarial network (GAN) setting to 3D, using volumetric convolutions to create a voxel-based generative model.
A followup by \citet{liu2017interactive} allows users a finer granularity of control by interactively painting a voxel grid and converting it to a high-resolution shape using a GAN. \citet{Girdhar16b} leverage a shared latent space of 2D views and voxels to generate shapes conditioned on images.
These ideas were later expanded beyond voxels to the non-Euclidean domains of point clouds \cite{li2018point, yang2019pointflow} and meshes \cite{ranjan2018faces, Tan2018VariationalAF, nash2020polygen}.
See \citet{Chaudhuri2019LearningGM} for a contemporary introduction to the field.

\textit{3D part-level representation.} Decomposing 3D shapes into parts was traditionally proposed as means of improving shape representation, usually to facilitate downstream tasks involving recognition, retrieval or manipulation \cite{hoffman1984parts, mitra2014structure, Huang2014FunctionalMN}.
With the rising popularity of data-driven approaches \cite{Kalogerakis2010learning, kim2013learningpartbased}, neural architectures have also been augmented with part level segmentation as a means of enriching shape representations \cite{qi2016pointnet, dgcnn2019}. 

An emerging trend promotes the encoding of shape parts in a joint latent space to facilitate better generalization of the representation to novel, unseen shapes \cite{nash2017shapevae}.
Other works attempt to achieve this goal by addressing the geometry and structural composition of parts separately \cite{lin2019sdmnet}.  Although these approaches are able to capture fine geometric details, they require part supervision. 
To avoid the need for labels, shape parts can be composed as deep hierarchies using binary-space partitions \cite{chen2020bspnet}, recursive neural networks  \cite{li2017grass, li2019grains, Paschalidou2020CVPR}, and Gaussian mixture models (GMM) \cite{achlioptas2018learning, Hertz_2020_CVPR}.
Our work uses a part decomposition network that is close in spirit to \cite{Hertz_2020_CVPR}, but we opt for a simplified variant of a flat GMM, rather than a hierarchy.

\textit{Neural implicit shapes.}
Non-neural implicit representations have been developed and used in a variety of 3D applications, such as reconstruction, modeling and morphing  \cite{10.1145/274363.274366, 10.1145/383259.383266,10.1145/1198555.1198639, schmidt2011shapeshop}. In these classical works, 3D shapes are implicitly approximated through a pre-defined family of functions, or interpolated distance fields. With the advances of learning based techniques, coordinate-based neural networks gain attention as powerful parameterizations able to fit arbitrary signals, and in particular, implicit shapes. 
Neural implicit functions are used to capture the geometry of 3D shapes as occupancy indicator functions \cite{chen2019learning,Mescheder2019occupancy, peng2020convolutional}, level sets of distance fields \cite{park2019deepsdf, Atzmon_2020_CVPR}, or indirectly as volumetric radiance fields \cite{mildenhall2020nerf, kaizhang2020nerfplusplus}.
Pioneering works by \citet{chen2019learning}, \citet{Mescheder2019occupancy} and \citet{park2019deepsdf} show how multiple shapes can be decoded with a single network by encoding shapes as latent codes. They are able to achieve high-quality shape reconstruction with a continuous representation that learns priors from a 3D dataset.

Neural implicit representations have been expanded to hybrid representations based on spatial structures of latent codes. \citet{peng2020convolutional}, \citet{jian2020localimplicitgrid} and \citet{chabra2020deepls} promote the usage of local dense grids as a means of introducing an inductive bias of spatial repetitions. \citet{martel2021acorn} and \citet{takikawa2021neural} use hierarchical octree representations to achieve faster rendering and higher reconstruction quality. 
Despite their promising results, directly applying these methods to shape generation and editing is non-trivial.

Various works advocate sparse representations where part templates are decoded with neural implicit functions. \citet{yang2018foldingnet} and \citet{groueix2018atlasnet} learn 3D shape representations as surface elements parameterized by 2D to 3D coordinate-based mapping networks. Closer to our work, SIF \cite{Genova2019LearningST} and LDIF \cite{genova2020local} study  representations that decompose shapes into coarse Gaussian template parameterizations, further localized with implicit surface functions to obtain a full shape reconstruction. A key observation is that the usage of template parts allows for smooth interpolation in latent space between shapes, suggesting that implicit part templates are promising for geometric editing. Unlike these works, we avoid the usage of encoders in favor of a more direct method to achieve decomposition into Gaussian parts. In \ourmethod, we also augment part representations with contextual information to achieve global coherency, suitable for shape editing. See \citet{xie2021neural} for summary of latest achievements in this field.

\textit{Interactive editing.} In this paper, we refer to editing as manipulation of non-atomic 3D shape elements by means of mixing parts \cite{funkhouser2004modeling} or through guided transformations, which leverage prior knowledge about the shape structure \cite{gal2009iwires,fakscm_metarep_sig14, mitra2014structure}.
Prominent examples of neural modifiers include editing of complex shapes through learned approximations of coarse primitives \cite{abstractionTulsiani17, hao2020dualsdf}, deformation networks \cite{wang20193dn, Yifan:NeuralCage:2020, jiang2020shapeflow, uy-joint-cvpr21} or segmenting and manipulating semantic parts \cite{9320459}. To extend these methods to new shapes, unseen during training, some degree of shape encoding or inversion is required.

Neural solutions for mixing parts have been investigated in MRGAN \cite{gal_iccvw21} and SP-GAN \cite{li2021spgan} for point clouds, which also allows for part-generation and part-mixing. Pertaining coordinate-based networks, COALESCE \cite{yin2020coalesce} studies stitching of parts through the synthesis of joint connections between them. Nonetheless, COALESCE requires segmentation labels. 

Most of the aforementioned methods do not allow part-level mixing or interpolation between neural implicit shapes.
Our method is the first to allow for both types of editing operations seamlessly on neural implicit surfaces. In Table \ref{tab:app}, we summarize the properties and applications of notable coordinate based methods.
Among these methods, DualSDF \cite{hao2020dualsdf} is the closest to ours. DualSDF learns mirrored coarse and fine representations per shape, using primitives and an implicit signed distance function, respectively. Their interaction goal is different from ours: they perform shape manipulation by minimizing the objective function over transformed primitive attributes. Since some of these attributes are unconstrained, primitives that are not directly manipulated by the user may still be affected by the optimization. Thus, the method cannot guarantee that global shape attributes are maintained during editing (Figure \ref{fig:local_edit}).
In contrast, our method uses sparse, learned representations that allow for direct editing of local shape parts while remaining faithful to the global shape structure. Since our method is part-aware, it also naturally supports mixing of parts from other shapes, as well as unconditional generation of novel shapes.

\newcommand{\cmark}{\ding{51}}
\newcommand{\cmarkb}{\ding{52}}
\newcommand{\xmark}{\ding{55}}

\newcommand{\cxmark}{\stackengine{0pt}{\cmark}{\xmark}{O}{c}{F}{T}{L}}
\begin{table}
\centering
\caption{Neural implicit shapes methods and applications.}
\label{tab:app}
\begin{tabular}{l \ts c \ts c \ts c \ts c}
\toprule
{Method} & {Inversion} & {Generation} & {Editing}  & {Mixing}\\
\midrule
DeepSDF \shortcite{park2019deepsdf}   & \cmark & \cmark  &  \xmark & \xmark\\
IM-NET\shortcite{chen2019learning}  & \cmark & \cmark  &  \xmark & \xmark\\
OccNet \shortcite{Mescheder2019occupancy}  & \cmark & \xmark  &  \xmark & \xmark\\
LDIF \shortcite{genova2020local} & \cmark & \xmark  &  \xmark & \xmark\\
COALESCE \shortcite{yin2020coalesce} & \xmark & \xmark  & \xmark & \cmark\\
DualSDF \shortcite{hao2020dualsdf} & \cmark & \cmark  &  \cmark & \xmark\\
\ourmethod{} & \cmark & \cmark  &  \cmark &  \cmark\\
\bottomrule
\end{tabular}

\end{table}

\begin{figure}
\centering
\includegraphics[width=\columnwidth]{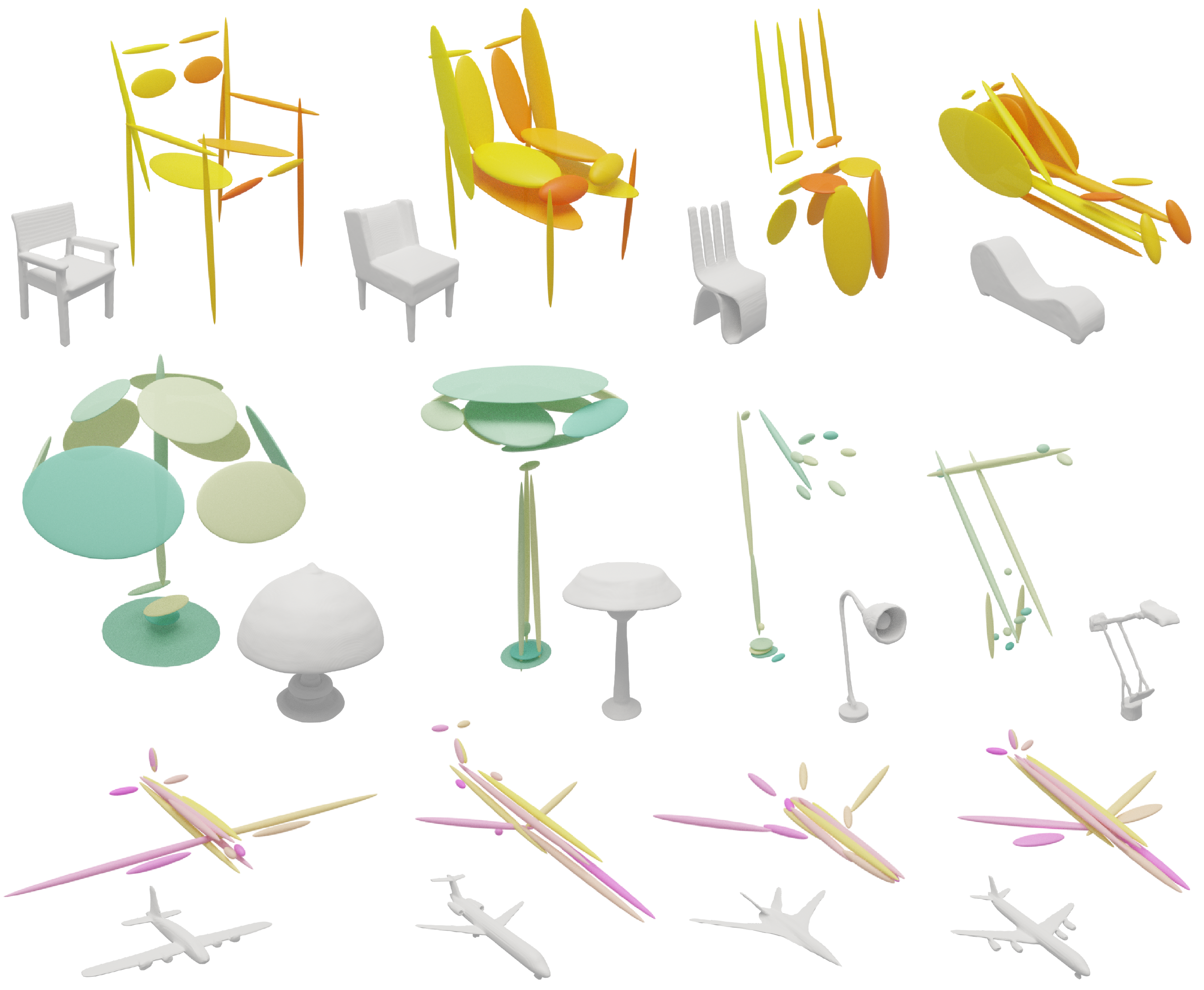}
\caption{Decomposition level. \ourmethod{} represents each learned shape as a GMM (colored blobs) and composes them into an implicit shape (in grey).}

\label{fig:gmms}
    
\end{figure}
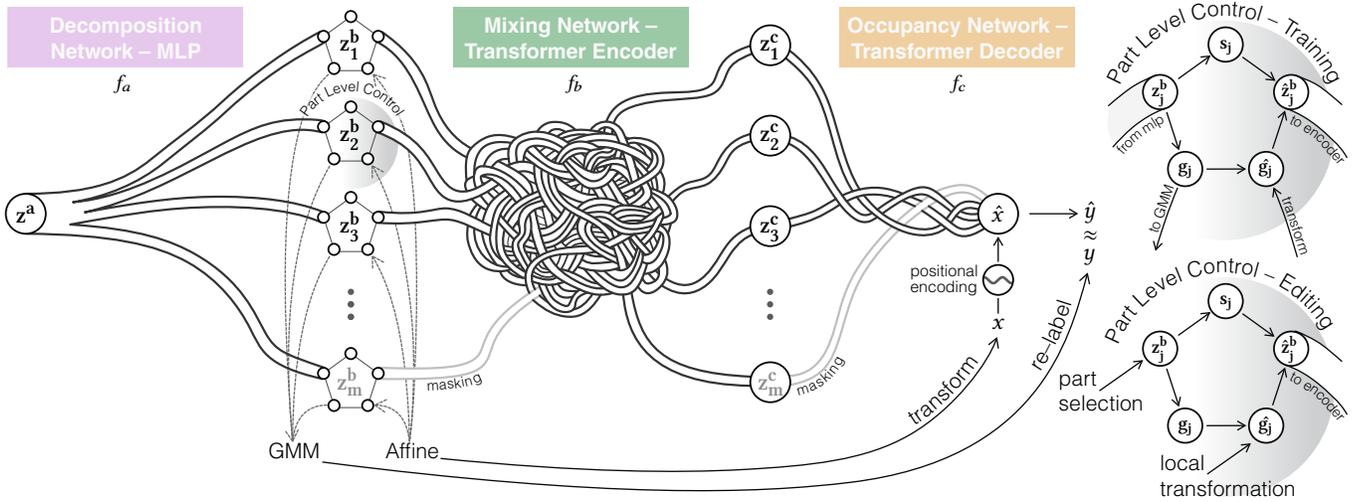
\begin{figure*}
\centering
    \begin{overpic}[width=1\textwidth,tics=10, trim=0 0 0 0,clip]{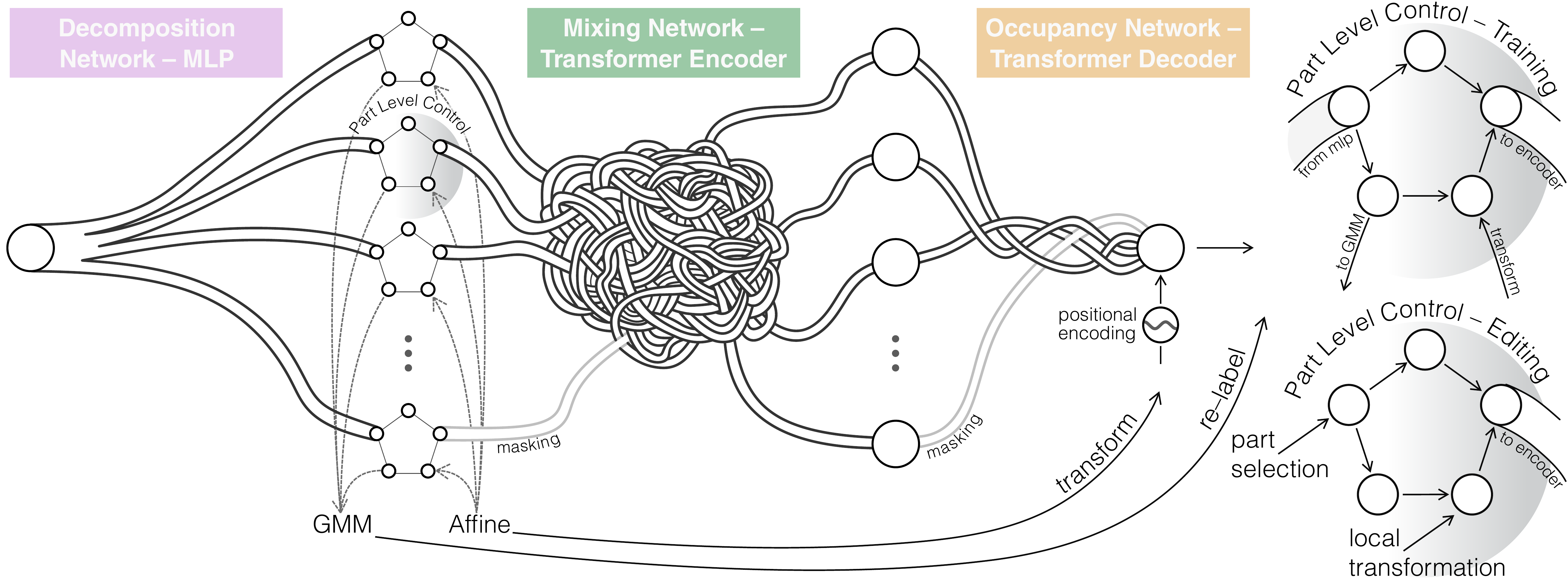}
  
     \put(1.2,20.8){$\mathbf{\za}$}
     \put(25.2,33.5){$\mathbf{\zb_{1}}$}
     \put(25.2,26.8){$\mathbf{\zb_2}$}
     \put(25.2,20){{$\mathbf{\zb_3}$}}
     \put(25,8.4){{\gray{$\mathbf{\zb_m}$}}}
      
     \put(56.35,33.5){$\mathbf{\zc_{1}}$}
     \put(56.35,26.8){$\mathbf{\zc_2}$}
     \put(56.35,20){{$\mathbf{\zc_3}$}}
     \put(56.2,8.4){{\gray{$\mathbf{\zc_m}$}}}
    
    \put(73.6,12.9){{$x$}}
    \put(73.55,20.85){{$\hat{x}$}}
    \put(80.4,21.1){{$\hat{y}$}}
     \put(80.3,19.35){$\approx$}
    \put(80.35,18){{$y$}}
    
    \footnotesize
    \put(85.4,30.1){$\mathbf{\zb_j}$}
    \put(87.35,24.5){{$\mathbf{g_j}$}}
    \put(93.3,24.5){{$\mathbf{\hat{g_j}}$}}
    \put(90.45,33.7){{$\mathbf{s_j}$}}
    \put(95,30.1){{$\mathbf{\hat{z}^{b}_{j}}$}}
    
    \put(85.4,11.1){$\mathbf{\zb_j}$}
    \put(87.35,5.4){{$\mathbf{g_j}$}}
    \put(93.3,5.4){{$\mathbf{\hat{g_j}}$}}
    \put(90.45,14.7){{$\mathbf{s_j}$}}
    \put(95,11.1){{$\mathbf{\hat{z}^{b}_{j}}$}}
     
     \put(8.5,30.7){$\netA$}
     \put(42,30.7){$\netB$}
     \put(70.5,30.7){$\netC$}

    \end{overpic}

    \begin{minipage}[t]{0.73\textwidth} 
        \subcaption[first caption.]{\ourmethod's pipeline. We train a \textit{decoder only} network that (i) decomposes a shape embedding ${\za}$ into distinct parts embeddings ${\zb_j}$ that correspond to a Gaussian mixture model (GMM); (ii) processes the (masked) set of ${\zb_j}$ into contextual embeddings ${\zc_j}$ and (iii) outputs an implicit shape, where a query coordinate $\xcoord$ is projected into a high-dimensional space using positional encoding and is then weighted according to $\zC$ in order to determine the occupancy indicator $\hat{y}$.
        The part disentanglement is achieved using the self-supervision, provided by the GMM for re-labeling the ground truth labels $y$. The local transformation control is achieved by applying rigid transformations on the both the Gaussians and the query coordinates.}
    \label{fig:method_diagram_a} 
    \end{minipage}\hspace{3mm}\begin{minipage}[t]{0.24\textwidth}
    \subcaption{Part level controller. Our network is trained (top) to have a disentangled representation of the surface geometry ($\sj$) and its rigid attributes ($\hat{\mathbf{g}_j}$). During inference (bottom),  the user can compose and manipulate the shape parts.}
    
    \label{fig:method_diagram_b} 
    \end{minipage}
    \caption{Method overview. Left: the network architecture. Right: a zoom-in into the part level control component. }

    \label{fig:method_diagram}
    
\end{figure*}

\section{Method}

Our goal is to establish a framework for editing 3D shape parts represented as learned neural implicit fields.
There are three key challenges which guide our design.

First, as part-level labels are expensive to obtain, we would like to learn the actual part decomposition during training. Specifically, we require the parts we form to be compact but sufficiently descriptive, as over-fragmentation may complicate the interactive editing process, and under-fragmentation may limit the degrees of freedom it allows.

Second, the latent part representations need to be disentangled from each other. This is essential to allow individual part editing and shape mixing. At the same time, the aggregation of parts should form a globally coherent 3D shape.

Third, the mapping between the latent representation and their corresponding occupancy indicators should ideally be equivariant with respect to affine transformations. Such a design allows us to directly manipulate the latent shape representation of parts during editing, while visually mirroring these transformation results in 3D space ad-hoc.

These requirements form the foundation of our architecture. In our formulation, shapes are represented and stored by a single learned global code. Our framework comprises of three parts, as summarized in the high-level overview of Figure \ref{fig:overview}.
Note, in the following sections, $\za, \zb$ or $\zc$ denote a high dimensional vector; $\zA, \zB$ or $\zC$ denote a collection of such vectors.

The first module, the Decomposition network $\netA$, maps the shape code $\za$ to a sparse representation of $m$ parts denoted as $\zB$. We fix $m$ to roughly agree with the cardinality of distinct shape parts. Each of these representations is projected to the parameters of a 3D Gaussian. There are numerous advantages to this choice. Predominately, it allows us to train shape representation $\za$ as a Gaussian Mixture Model, which naturally forces our architecture to maintain global shape coherency. Likewise, it allows our network to leverage priors about shape categories, as Gaussians are implicitly encouraged to fit parts which are common across all shapes. Most importantly, this decomposition allows for direct control over the shape parts, by applying 3D transformations on the Gaussians associated with them.

The second component, Mixing Network $\netB$, augments the part representations with contextual information, which further ensures that the local part embeddings remain aware of the global shape structure.

The final building block, Occupancy Network $\netC$, decodes the contextual embeddings to binary occupancy values, essentially forming our neural implicit representation.

Note that we treat our end-to-end framework as an auto-decoder which jointly trains an implicit shape generative model and embedding vectors $\zA = \set{\za}{i}{n}$ corresponding to $n$ training examples. To simplify the notations, we omit the shape index $i$ for the rest of this section. In practice, we train over datasets of shapes from the same class; see further training details in \ref{sec:imp}.  

The full architecture of \ourmethod{} is detailed in Figure~\ref{fig:method_diagram}, and accompanies the in-depth discussions for the rest of this chapter.

\subsection{Decomposition network}
\label{sec:decomposition}

Our first objective is to obtain a part-level decomposition of a given shape, which forms the basis of our part based editing. Previous works \cite{genova2019learning, Chen_2019_ICCV, Hertz_2020_CVPR} have shown the ability of 3D generative neural networks to learn a consistent part-level decomposition across generated shapes, without using explicit part-level supervision. Similarly, \ourmethod{} utilizes this ability by conditioning the shape generation on the parts partitioning, and their manipulation.

We formulate the Decomposition network $\netA$ as a decoder, trained to map input shape code $\za$ into a GMM representation. Given a shape embedding $\za$, we first split it into $m$ distinct vectors $\netA(\za) = \zB$, where $\zB \in \reals^{m \times \dm}$ is a set of high dimensional parts embeddings $\set{\zb{}}{j}{m}$. The encoding of each part $\zb{}_j \in \dm$ is further projected to two sets of parameters: intrinsic surface geometry information $\sj \in \ds$, and extrinsic parameters represented by the Gaussian $\gj \in \reals^{16}$ (see Figure~\ref{fig:method_diagram_b}, top):
\begin{equation}
\begin{aligned}
\label{eq:project}
& \sj = W_s \zb{}_{j} + \mathbf{b}_s \\
& \gj = W_{d} \zb{}_{j} + \mathbf{b}_{d}
\end{aligned}
\end{equation}
Intuitively, $\gj$ marks the area of influence of each part $j$, whose detailed structural information is captured by $\sj$. One of the advantages of this representation is that across the entire dataset, similar intra-category parts are represented using the same Gaussians in a consistent way.

The decomposition network $\netA$, is a multi-layer perceptron (MLP) where after the first fully connected layer, we split the embedding to $m$ vectors and the rest of the layers are shared between the $m$ embeddings. It is  followed by the projection of $\zb{}_{j}$ onto the pair $\sj$ and $\gj$ (Eq~\eqref{eq:project}). See Appendix \ref{sec:imp_net} for further implementation details. 

The low dimensional Gaussian $\gj$ is a stacked representation of the parameters: mixing weight $\pi_{j} \in \reals^{1}$, center $\boldsymbol{\mu}_{j} \in \reals^3$ and factorized covariance matrix values $U_{j} \in \reals^{3 \times 3}$, $\boldsymbol{\lambda}_{j} \in \reals^{3}$. The covariance matrix can be calculated using the eigendecomposition $\Sigma_{j} = U_{j}^{-1} D_{j} U_{j}$, where $D_{j}$ is a diagonal matrix with the vector $\boldsymbol{\lambda}_{j}$ as its diagonal and $U_{j}$ is a unitary matrix.

Given a batch of points $\xin \in \reals^{B \times 3}$ randomly sampled \emph{inside} the full shape that corresponds to the global embedding $\za$, the network
$\netA$ is trained by the GMM negative log-likelihood loss:
\begin{equation}
\mathcal{L}_{GMM}(\xin, GMM) = -\log p \left(\xin \vert \textrm{GMM}\right),
    \label{eq:loss_gmm}
\end{equation}
where:
$$
p \left(\xin \vert \textrm{GMM}\right) = \prod_{\mathbf{x} \in \xin}\sum_{j=1}^{m} \pi_{j} \mathcal{N}\left(\mathbf{x} \vert \boldsymbol{\mu}_{j}, \Sigma_{j} \right). 
$$

The GMM loss encourages the decomposition network $\netA$ to dissipate the Gaussians over the entire shape volume, such that every randomly sampled point can be explained by at least one of the Gaussians in the mixture.
Figure~\ref{fig:gmms} shows examples for the GMM decomposition learned for various shapes.

\subsection{Implicit shape composer}
\label{sec:composition}

The rest of our network is trained to compose together the set of part embeddings $\zB{}$, to a high resolution implicit shape. 
Each part representation $\zb \in \zB{}$ from the Decomposition network is disentangled to extrinsic and intrinsic components, which are then reconstructed back together to form representations $\zbh\in\zBh$. For brevity, we defer the discussion about attribute disentanglement to Section~\ref{sec:att_disentanglement}, and resume directly from where $\zbh$ are fed to the rest of the pipeline (Figure~\ref{fig:method_diagram_b}).

Our composition begins by using the Mixing network $\netB$ to reinforce the representation of each part $\zbh{}$ with contextual information from other parts in $\zBh{}$. The Mixing network outputs part representations $\zC = \netB(\zBh)$, that are global aware.
Then, we use Occupancy network $\netC$ to obtain the composed implicit function $\hat{y}$. Given a query coordinate $\xcoord \in \reals^3$, we calculate $\hat{y}=\netC(\xcoord| \zC)$, e.g: attend coordinate $x$ on contextual representations $\zC$. This yields the part-aware coordinate embedding $\xcoordhat$, which we then proceed to decode for its occupancy value $\hat{y}$.

Both networks are realized through the full Transformer architecture of \citet{vaswani2017attention} where Mixing network $\netB$ is realized through the Transformer encoder, and $\netC$ is a customized variant of the decoder. The Transformer is suitable for our task due to its powerful capabilities of learning contextual representations from sequences or unordered sets in various domains \cite{radford2018improving, devlin2018bert, lee2019set, zhao2021point}.

The Transfromer encoder of Mixing network $\netB$ does not use positional encoding, since we're interested in embeddings of an unordered set. Global aware representations $\zC$ are obtained through a series of multi head attention layers: 

\begin{equation}
\textrm{Attention}(Q, K, V) = \textrm{softmax}\left(\dfrac{QK^T}{\sqrt{d_k}} \right) V.
\label{eq:attention}
\end{equation}

Under the formulation of \citet{vaswani2017attention}, the queries, keys and values matrices of \emph{each} head $h$ in layer $t$ are given by projecting respectively:
\begin{align*}
Q_e = \zBh W^{Q_e}; && K_e = \zBh W^{K_e}; && V_e = \zBh W^{V_e}  \\
W^{Q_e} \in \reals^{\dm \times d_k}; &&  W^{K_e} \in \reals^{\dm \times d_k}; &&  W^{V_e} \in \reals^{\dm \times d_v}
\end{align*}
with dimensions $d_k = d_v = \dm / h$. 
The final embedding of each shape part, $\zc \in \dm$, is obtained by concatenating the per-head outputs and projecting them together.

The Transformer decoder of Occupancy network $\netC$ uses the output from $\netB$ and coordinates $\xcoord \in \reals^3$. During training, coordinates $\xcoord$ are sampled along with their shape occupancy label $y$ around the surface and within a bounding volume $[-1, 1]^3$. We denote each batch of sampled pairs as
$\xbatch=\big\{ (\xcoord{_i}, y_i) \big\} ^{B}_{i=1}$. Appendix \ref{sec:imp_data} for further elaborates about the sampling scheme and data preparation.

Before feeding coordinates $\xcoord$ to the decoder attention blocks, we first project them onto a high dimensional space $PE(\xcoord) \in \reals^{\dpe}$ using a $learned$ positional encoding layer. To avoid potential ambiguity, we clarify that positional encodings were previously mentioned in the context of Transformers as means of preserving order in sequences. Similar formulations have been discussed in the literature of Neural Implicit Fields \cite{tancik2020fourfeat} as means of increasing the network sensitivity to coordinate based input and overcoming \emph{Spectral Bias} \cite{rahaman2019spectralbias}. Our formulation refers to the latter definition, pertaining coordinate based networks, and is closer in definition to a single SIREN layer \cite{sitzmann2019siren}:
\begin{equation}
\textrm{PE}(\xcoord) = \sin{\left(a (W_{pe}\xcoord + B_{pe})\right)},
\end{equation}
where $W_{pe} \in \reals^{\dpe \times 3}$ and $B_{pe} \in \reals^{\dpe}$ are learned parameters and $a$ is a fixed scalar. Using a learned variant allows us for an easier initialization which avoids careful tuning due to scale sensitivity issues common in deterministic PE parameterizations.
\cite{hertz2021sape}.

Network $\netC$ proceeds to calculate the part-aware coordinate embedding $\xcoordhat$ with a sequence of $T$ cross attention layers (Eq.~\eqref{eq:attention}), enumerated as $0 \leq t <T$. From a \emph{single coordinate} point of view, layer $t$ in  $\netC$ outputs the embedding $\xcoordhat_{t+1}$, calculated by the cross attention of $\xcoordhat_{t}$ with $\zC$:
\begin{equation}
\begin{aligned}
\textbf{q}_d = \xcoordhat_{t} W^{Q_d}; && K_d = \zC W^{K_d}; && V_d = \zC W^{V_d} \\
W^{Q_d} \in \reals^{\dpe \times d_k}; &&  W^{K_d} \in \reals^{\dm \times d_k}; &&  W^{V_d} \in \reals^{\dm \times \dpe}
\end{aligned}
\label{eq:cross_terms}
\end{equation}
We define $\xcoordhat_{0} = \textrm{PE}(\xcoord)$, and designate the final output as $\xcoordhat = \xcoordhat_{T}$. Each attention layer consists of multi-head attention followed by a position-wise feed-forward network. 

Unlike the \textit{classic} transformer decoder, we omit the self-attention layers from the decoder. This is essential for a couple of reasons: (i) The occupancy indicator of a coordinate $\xcoord \in \xbatch$ should be agnostic to other coordinates we feed in the same set $\xbatch$ with $\xcoord$, and (ii) We assume the amount of sampled points is considerably larger than the number of parts, e.g: $B >> m$. While it is acceptable to allow quadratic dependency on $m$, it is desirable to keep the network run-time complexity linear in $B$. 

The last part of $\netC$ is a MLP which decodes $\xcoordhat$ to yield an occupancy indicator $\hat{y}$.
The occupancy loss is given by the binary cross entropy loss ($BCE$):

\begin{equation}
\mathcal{L}_{occ}\left(\xbatch, \zBh\right) =  \dfrac{1}{|\xbatch|} \sum_{\left(x,y\right)\in \xbatch}BCE \left(\hat{y}, y \right),
    \label{eq:loss_occ}
\end{equation}

Notice that the contextual-part $\zC$ is indifferent to the query coordinates. Therefore, in order to reconstruct a $3D$ shape from $\za$, we feed forward the mapping network and the transformer encoder once. Then, the decoder operates in parallel for multiple coordinates, while $\zC$ remains fixed.

\subsection{Disentanglement of extrinsic attributes}
\label{sec:att_disentanglement}

We turn to introduce the extrinsic-geometry disentanglement component, which enables local transformation control over the generated shapes (Figure~\ref{fig:method_diagram_b}).
Recall, we have already retrieved the geometric properties for each part representation, $\zb{}_{j}$, by projecting it to the stacked representation of Gaussian $\gj$ (see Section~\ref{sec:decomposition}). In addition, we obtained the detailed surface information representation $\sj$.

In the following, we attempt to make embedding $\sj$, \emph{invariant} to affine transformations applied over the shape part. At the same time, we want the mapping of $\gj$ to the shape part geometry to be \emph{equivariant} with respect to affine transformations. In other words, transformations applied on $\gj$ should be directly mapped to the decoded shape part $j$.

To that end, we apply a random affine transformation $\mathit{T}$ on $\gj$ to obtain the transformed Gaussian $\gjhat$. We then up-project and inject it back to $\zb{}_{j}$ by:

\begin{equation}
\zbh{}_{j} =  \underbrace{\sj}_{\textrm{intrinsic}} + \ \ \  \underbrace{W_{u} \gjhat + \mathbf{b}_{u}}_{\textrm{extrinsic}}.
\end{equation}

The modified set $\hat{\zB} = \set{\zbh{}}{j}{m}$ is then routed to the composition networks as described in Section~\ref{sec:composition}. Finally, we apply the same transformation $\mathit{T}$ on $\xbatch$ such that output implicit function $\hat{y}$ learns to mimic the transformed shape. 

On one hand, any extrinsic attributes, i.e., part location and orientation, that might be concealed in $\sj$ are now irrelevant for the reconstruction of $\hat{y}$. This is true since transformation $\mathit{T}$ is applied only on $\gj$. On the other hand, $\gjhat$ does not contain any intrinsic surface geometry information by construction. It is extracted from the low dimensional embedding $\gj$, which holds the parameters of a single Gaussian. Thus, disentanglement of extrinsic and intrinsic geometric attributes is achieved.

\subsection{Part-level disentanglement}
Ideally, we would like part representations $\zB$ to contain only local part information, and contextual representations $\zC$ to be global-aware. The former is required to allow an intuitive part-editing mechanism, where the latter is crucial to have a high-quality reconstruction of the shape $\hat{y}$.

 Even though, each part embedding ${\zb{}_{j}}$ corresponds to a single shape Gaussian, there is no guarantee that additional global information will not \textit{leak} between different part embeddings of the same shape.
Indeed, such ``leaks'' may harm the quality of local editing and our ability to mix parts between shapes (see Section~\ref{sec:ablation}).

To overcome this problem, we augment each training iteration with an additional forward pass, which promotes $\zB$ to contain local information and better separate the Gaussians area of effect. Different to before, we carefully select a subset of part embeddings, denoted as $Z^{b_-} \subset \zB$. Specifically, we're interested in choosing part representations that may contain mutual knowledge about each other, which is more common with Gaussians that are proximate or potentially overlap.
We therefore randomize a direction vector $\mathbf{u} \in \mathbb{R}^3$, and sort all Gaussian centers in that direction. Then, we choose $L$ sequential Gaussians whose part representations constitute $Z^{b_-}$.

From here, we proceed to reconstruct $\hat{y}$ as usual. We expect that the output implicit shape of this forward pass will contain \emph{only} shape parts that are governed by representations $Z^{b_-}$.

Since we do not have direct supervision to guide this optimization, we utilize the clustering induced by the $GMM$ of the sampled points $\xbatch$, to generate self-supervised labels. We assign each coordinate $\xcoord \in \xbatch$ to the Gaussian that maximize its expectation:
$$
Cluster \left(\xcoord |GMM \right) = \argmax_j \pi_j \mathcal{N}\left(\xcoord \vert \boldsymbol{\mu}_{j}, \Sigma_{j} \right),
$$ then, we re-label occupancy $y$ as:
\begin{equation}
\label{eq:re-label}
    y^{-}=\begin{cases}
    y,& \text{if } Cluster \left(\xcoord | GMM \right) \in \mathbb{I} (Z^{b_-})\\
    0,              & \text{otherwise,}
\end{cases}
\end{equation}
where $\mathbb{I}(Z^{b_-})$ are the indices of the Gaussians in $Z^{b_-}$. Intuitively, we relabel the \emph{outside-coordinates}, e.g: the coordinates that are clustered outside the Gaussians of $Z^{b_-}$, as "non-occupied".
The part-level disentanglement loss is then given by an occupancy loss term calculated over the modified set:
\begin{equation}
\label{eq:loss_dis}
\mathcal{L}_{dis}(\xbatch, Z^{b_-}, GMM) = \mathcal{L}_{occ}(\xbatch^\textrm{-}, Z^{b_-}),
\end{equation}
where $\xbatch^\textrm{-}$ is the set of re-labeled pairs $(x, y^{-})$.

In practice, for parallel batch training, the selection of subset $Z^{b_-}$ is achieved by using attention masking. The attention weights of part embeddings not in $Z^{b_-}$ are forced to be zero. This is illustrated by the light gray straws in Figure~\ref{fig:method_diagram_a}.

\subsection{Training loss function}

The complete loss term for our network is given by the terms discussed so far:
\begin{equation}
\label{eq:loss_all}
   \mathcal{L}_{\ourmethod} = \mathcal{L}_{GMM} + \mathcal{L}_{occ} + \mathcal{L}_{dis} + \gamma \|\za\|_2,
\end{equation}
where $\gamma$ is a hyperparameter controlling the loss weight. We also apply regularization $\|{\za}\|_2$ per global shape embedding, as advocated by previous auto-decoder works \cite{bojanowski2018optimizing, park2019deepsdf}. The latent regularization promotes the shape codes to be normally distributed. That, in turn, makes the space of global shape codes easier to sample from.

\subsection{Shape inversion}
\label{sec:invert}
Our framework learns to represent shapes through an auto-decoder \cite{park2019deepsdf}.
To allow editing of a new shape that is not part of the training data, we first have to match it with a shape embedding $\za$. Our goal is to acquire part embeddings $\zB$ such that the generated implicit shape is \textit{as close as possible} to the new given shape. We suggest a simple two-steps optimization process to find these embeddings.

In the first step, we begin with a randomly initialized code: 
$$\za \sim \mathcal{N}\left(\boldsymbol{\mu}_{tr}, \Sigma_{tr} \right),
$$ 
using the mean and covariance of shape embeddings seen during training. We sample points $\xin$ inside the new given shape, and use $\netA$ with frozen weights to obtain $\zB$. Specifically, $\za$ is optimized using the GMM negative log-likelihood (Eq.~\eqref{eq:loss_gmm}) between the generated GMM of $\mathbf{\za}$ and sampled points, $\xin$, inside the new given shape.

In the second optimization step, we ensure the obtained $\zB$ reproduce the shape accurately. Therefore, we freeze the layers of $\netB$, $\netC$ and use occupancy loss (Eq. \eqref{eq:loss_occ}) to further optimize the part embeddings $\zB$:
$$
\argmax_{\zB} \mathcal{L}_{occ}(X_\textrm{cube}, \zC),
$$
where $\zC$ are given by the contextual encoder $\zC = \netB(\zB)$.

Upon convergence, we can use $\zB$ in our interactive interface as described in the following section.

\subsection{User interface for interactive shape editing} 
\label{sec:interactive}

We demonstrate how a pre-trained \ourmethod{} model can be used in an interactive editing framework. In this setting, the user can manipulate newly generated shapes, or reference shapes taken from an existing dataset. Examples of some editing operations can be seen in Figures~\ref{fig:teaser}, \ref{fig:col_mix} \ref{fig:local_edit}, as well as in the supplementary video.
Our coarse GMM representation is used for partitioning the generated shapes. The partition provided by the GMMs enables a simple interface for quick selection of shape parts. 

Users can select parts from different shapes, mix them together, and assemble new shapes. Under the hood, each part selection corresponds to selection of latent vectors from $\zB$. These selected latents can then be combined with latent vectors of selected parts from other shapes, to form a single set of latent codes. The combined latents are forwarded through our pretrained network (as shown in Figure~\ref{fig:method_diagram_a}) to synthesize newly mixed shapes.

In addition, users can select shape parts and apply affine transformations to them, which results with local deformation of the selected parts. As shown in Figure~\ref{fig:method_diagram_a}, the specified transformations are applied on the Gaussian parameters of the selected part. After transformations are applied, the Gaussian representation is injected back to the part latent representation, which goes to the transformer part of our network to synthesize the new manipulated shape.

Finally, we remark on the rasterization pipeline we used for rendering implicit shapes. After each editing step, we reconstruct a mesh with the marching cube algorithm \cite{lorensen1987marching}, using a grid resolution of $256^3$. To reduce the number of queries through the network \cite{takikawa2021neural, hedman2021snerg}, we backed the grid implementation with our in-house implementation of an Octree. The acceleration structure enables an interactive rate of about two seconds between each editing step. Future applications may opt to avoid the mesh conversion step and render the implicit shape directly using ray-marching techniques.

\begin{figure}
\centering
\begin{overpic}[width=1\columnwidth,tics=10, trim=0mm 0 0mm 0,clip]{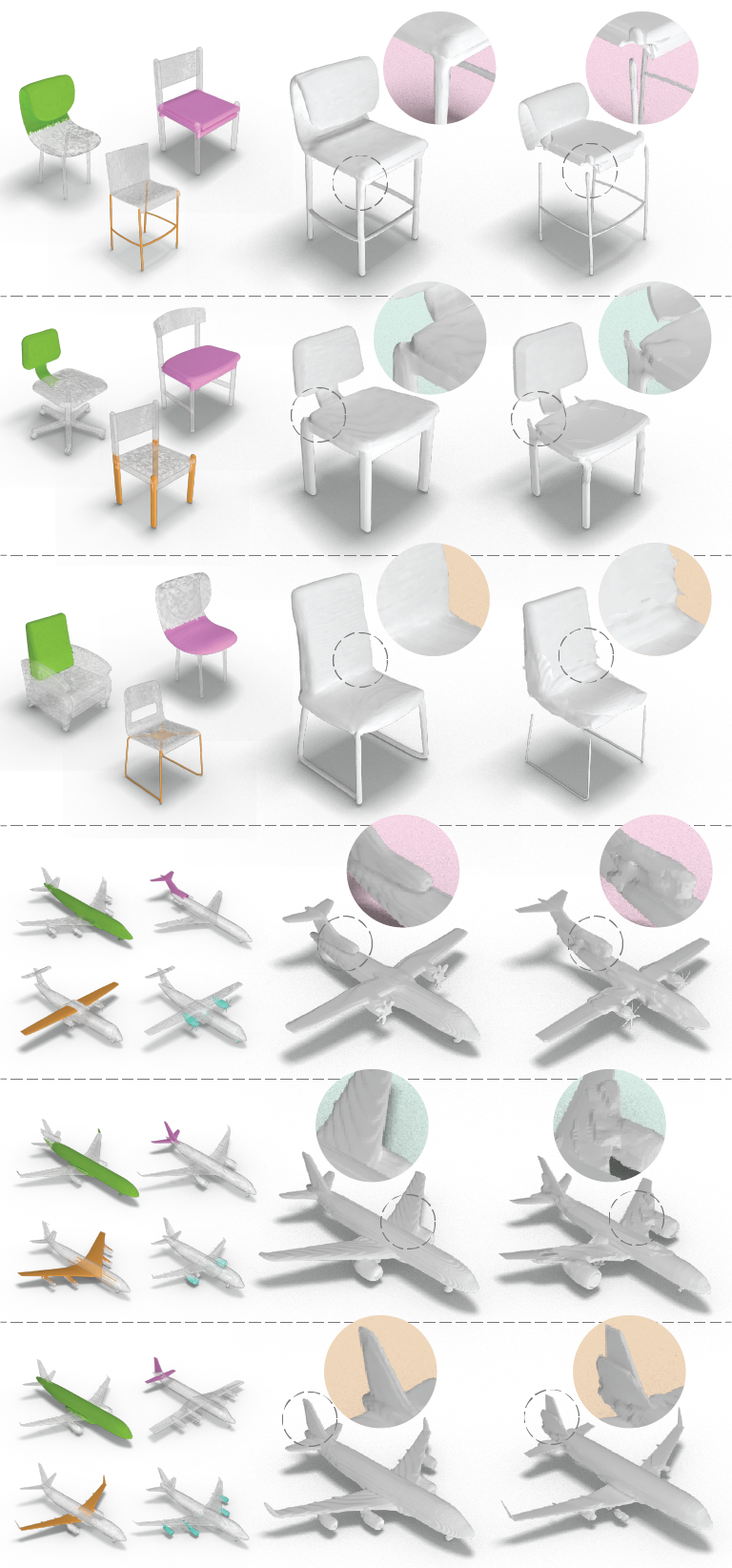}
    \put(5,-1){Input parts}
     \put(19,-1){SPAGHETTI}
    \put(35.5,-1){COALESCE}
    \end{overpic}

\caption{Mixing comparison. On the left are the input parts supplied to our method and COALESCE~\cite{yin2020coalesce} 
in order to create a unified novel \textit{mixed} shape. While COALESCE synthesizes only the joints between parts, our method synthesize the whole \textit{mixed} shape.}

\label{fig:col_mix}
    
\end{figure}
\begin{figure}
\centering
\footnotesize
\begin{overpic}[width=1\columnwidth,tics=10, trim=0mm 0 0mm 0,clip]{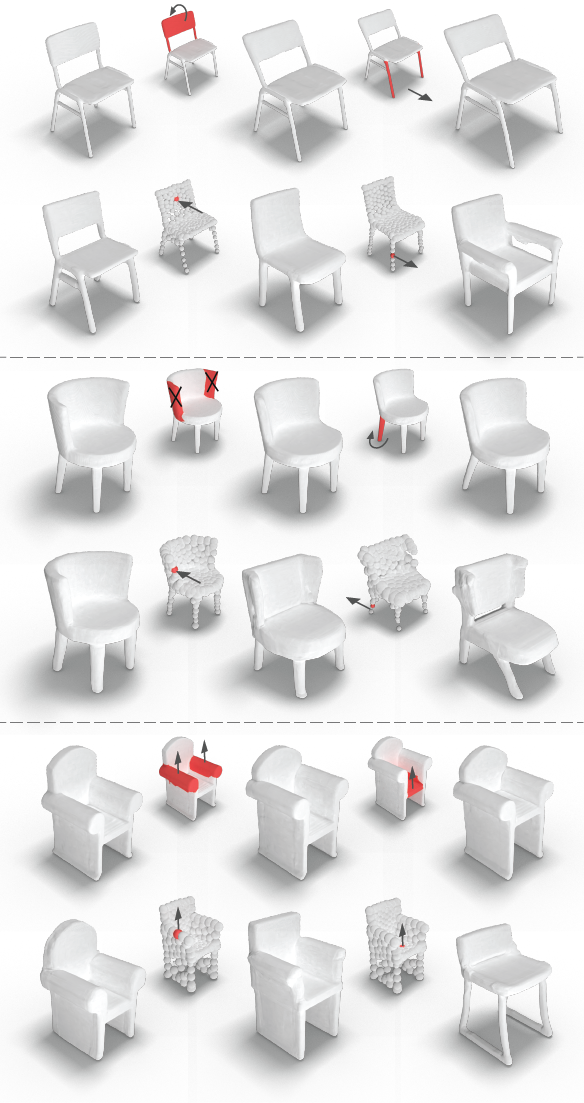}

     \put(1,84){\ourmethod{}}
     \put(1,69){DualSDF}
     \put(1,51.5){\ourmethod{}}
     \put(1,36){DualSDF}
     
    \put(1,18){\ourmethod{}}
    \put(1,2){DualSDF}
\end{overpic}

\caption{Sequential editing of implicit shapes using our method (top rows) and DualSDF \cite{hao2020dualsdf} (bottom rows). The input edit (marked in red) provided by the user is shown between the columns.}

\label{fig:local_edit}
    
\end{figure}
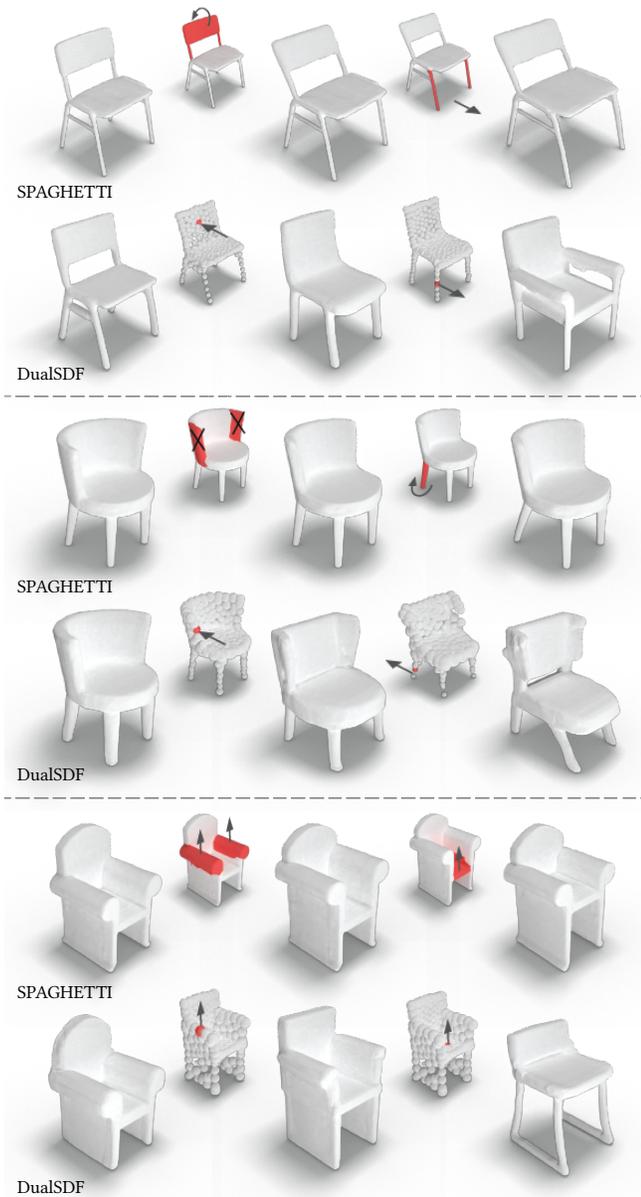

\begin{table*}
\centering

\caption{Shape inversion comparisons. CD = chamfer distance; EMD = earth mover’s distance; ACC: mesh accuracy \cite{seitz2006comparison}. In all measurements, lower score is better. All measurements are multiplied by a scale of $10^3$.}
\label{tab:inv}

\begin{tabular}{l \tss c \ts c \ts c  \tsl c \tss  c \tss c \tsl c \tss c \tss c}
\toprule

 {Method} & \multicolumn{3}{c \tsl \tsm}{{Airplanes}}  & \multicolumn{3}{c \tsl \tsm}{{Chairs}}  & \multicolumn{3}{c \tsl }{{Lamps}} \\ 

& $\text{CD}_{mean / med.}$ & $\text{EMD}_{mean / med.}$ & ACC 
& $\text{CD}_{mean / med.}$ & $\text{EMD}_{mean / med.}$ & ACC 
& $\text{CD}_{mean / med.}$ & $\text{EMD}_{mean / med.}$ & ACC \\

 \midrule
 IM-NET \shortcite{chen2019learning} &
 0.435 / 0.195 & 28.96 / 25.09 & 17.541
 & 0.625 / 0.436 & 33.40 / 30.53 & 22.062
 & 2.431 / 1.396 & 73.83 / 59.40 & 57.313\\

 LDIF \shortcite{genova2020local} & 0.634 / 0.178 & 31.50 / 20.23 & 17.237
 & 0.800 / 0.425 & 29.69 / 25.65 & 16.787
 & 3.076 / 0.684 & 56.18 / 42.36 & 45.478 \\
 DeepSDF \shortcite{park2019deepsdf} & 0.095 / 0.029 & 16.00 / 13.17 & 6.229 
 & 0.323 / 0.113 & 24.23 / 19.72 & 14.56 
 & 0.792 / 0.205 & 34.62 / 24.70 & 15.40 \\
 DualSDF \shortcite{hao2020dualsdf} & 0.806 / 0.097 & 31.93 / 22.20 & 26.78
 & 0.688 / 0.369 & 34.98 / 32.46 & 29.59
 & 2.906 / 0.827 & 75.04 / 50.12 & 46.69 \\
 \ourmethod{} & \textbf{0.050} / \textbf{0.011} & \textbf{9.27} / \textbf{7.17} & \textbf{4.237}
 & \textbf{0.102} / \textbf{0.032} & \textbf{13.55} / \textbf{11.26} & \textbf{6.884}
 & \textbf{0.559} / \textbf{0.041} & \textbf{17.05} / \textbf{12.17} & \textbf{6.671} \\
 \midrule
  \ourmethod{} no-enc & 0.052 / 0.014 & 10.84 / 8.66 & 4.599
  & 0.140 / 0.044 & 16.40 / 12.95 & 8.352
  & 0.966 / 0.061 & 22.20 / 16.41 & 7.410 \\
  \ourmethod{} no-dis & 0.068 / 0.015 & 11.09 / 8.43 & 5.310
  & 0.135 / 0.039 & 15.81 / 12.44 & 7.952
  & 0.854 / 0.083 & 23.81 / 16.52 & 11.30 \\

\bottomrule

\end{tabular}

\end{table*}

\begin{table*}
\centering

\caption{Shape generation comparisons. $\uparrow$ ($\downarrow$): higher (lower) is better. MMD-CD scores are multiplied by $10^3$; MMD-EMD scores are multiplied by $10^2$; JSD scores are multiplied by $10^2$. }
\label{tab:gen}

\begin{tabular}{l \tsl c \ts c \ts c  \ts c  \tsl c \ts c \ts c  \ts c \tsl c \ts c  \ts c \ts c}
\toprule

{Method}  & \multicolumn{4}{c \tsl \tsm}{{Airplanes}}  & \multicolumn{4}{c \tsl \tsm}{{Chairs}}   & \multicolumn{4}{c \tsl }{{Tables}} \\
& $\text{JSD}$\downa & {$\text{MMD}$ \downa} & {$\text{COV}$ \upa}  & {$\text{1-NNA}$ \downa}
& $\text{JSD}$\downa & {$\text{MMD}$ \downa} & {$\text{COV}$ \upa}  & {$\text{1-NNA}$ \downa}
& $\text{JSD}$\downa & {$\text{MMD}$ \downa} & {$\text{COV}$ \upa}  & {$\text{1-NNA}$ \downa}\\
\midrule
  && \cdemd& \cdemd & \cdemd && \cdemd & \cdemd & \cdemd && \cdemd & \cdemd &\cdemd \\

DeepSDF  & 3.89 & 3.8 / 10.2 & 32.6 / 33.5 & 70 / 71 & 1.62 & 11.1 / 13.6 & 41.2 / 44.3 & 60 / 61 & 1.35& 17.3 / 14.8 & 42.1 / 41.1 & 59 / 59 \\
IM-NET & 3.77 & 4.2 / 10.5 & 30.0 / 33.2 & 65 / 64 & 2.37 & 12.44 / 15.1 & 37.7 / 36.4 & 61 / 62 & 3.35& 16.6 / 15.7 & 37.7 / 39.2 & 61 / 61  \\
DualSDF   & 6.78 & 4.2 / 11.1 & 25.0 / 24.1 & 70 / 77 & 4.49 & 10.4 / 15.9 & 32.6 / 27.1 & 70 / 76 &  2.19& 12.3 / 15.2 & 36.3 / 32.7& 68 / 72  \\
\ourmethod{} & \textbf{2.28} & \textbf{2.4} / \textbf{8.10} & \textbf{35.0} / \textbf{41.3} & \textbf{61} / \textbf{61} & \textbf{1.02} & \textbf{6.01} / \textbf{11.4} & \textbf{50.8} / \textbf{51.2} & \textbf{58} / \textbf{59} & \textbf{1.15} & \textbf{5.9} / \textbf{11.1} & \textbf{47.8} / \textbf{48.8} & \textbf{56} / \textbf{56} \\
\bottomrule
\end{tabular}

\end{table*}

\begin{table}
\centering
\caption{\ah{Disentanglement ablation. Correspondence (Cor.) measures the IoU (\%) between distinct parts in our coarser GMM representation to the output implicit shape. Coverage (Cov.) measures the IoU of the whole output implicit shape and the union of its distinct parts. For both,  higher score is better.}}
\label{tab:abl_cor}

\begin{tabular}{l \ts c \ts c \tsm c \ts c  \tsm c \ts c}
\toprule

{Method} & \multicolumn{2}{c \tsm \ts}{{Tables}}  & \multicolumn{2}{c  \tsm  \ts}{{Chairs}}  & \multicolumn{2}{c \tsm}{{Lamps}} \\

& \ioupart & \iouall
& \ioupart & \iouall
& \ioupart & \iouall \\
 \midrule
 \ourmethod{} full & 0.779 & \textbf{0.857}
 & 0.670 & \textbf{0.765}
 & 0.747 & \textbf{0.854} \\
 
  \ourmethod{} no-enc & \textbf{0.810} & 0.829
  & \textbf{0.673} & 0.736
  & \textbf{0.795} & 0.781\\
  \ourmethod{} no-dis & 0.423 & 0.527
  & 0.259 & 0.411 
  & 0.455 & 0.661
  \\
\bottomrule

\end{tabular}

\end{table}

\begin{table}
\centering

\caption{\ah{Mixing ablation. Segmentation (Seg.) measures the Jensen-Shannon divergence (multiplied by $10^2$) between segmented parts of the generated mixed shapes to the ground truth input parts. Area measures the surface area error (\%) between them. For both, lower score is better.}}
\label{tab:abl_mix}

\begin{tabular}{l \ts c \ts c \tsm c \ts c  \tsm c \ts c}
\toprule

{Method} & \multicolumn{2}{c \tsm \ts}{{Tables}}  & \multicolumn{2}{c  \tsm  \ts}{{Chairs}}  & \multicolumn{2}{c \tsm}{{Lamps}} \\

& Seg. & Area
& Seg. & Area
& Seg. & Area \\
 \midrule
 \ourmethod{} full & \textbf{1.89} & \textbf{20.6}
 & \textbf{1.64} &  \textbf{9.9}
 & \textbf{7.03} & \textbf{12.3}\\
 
  \ourmethod{} no-enc & 2.39 & 23.1
  & 2.06 & 12.9
  & 7.55 & 14.0 \\
  \ourmethod{} no-dis & 3.33 & 29.2
  & 2.66 & 17.9
  & 9.92 & 25.1 \\
\bottomrule

\end{tabular}

\end{table}

\begin{figure}
\centering
\includegraphics[width=\columnwidth]{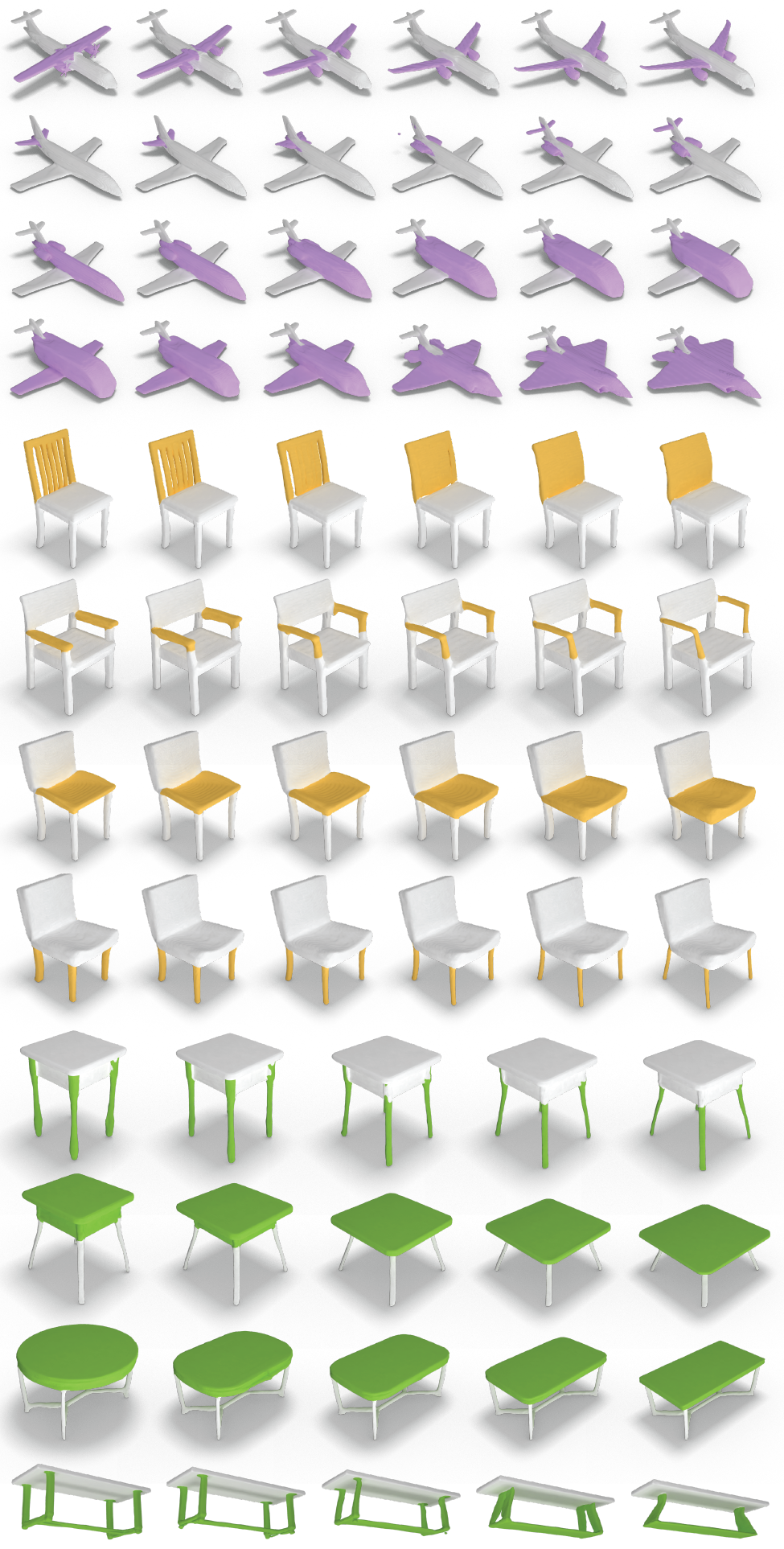}
\caption{Part level interpolation. By interpolating the attention weight of $\netC$, \ourmethod{} can make continuous interpolation between a selected specific part of a shape or any number of parts. In color are the modified parts.}
\label{fig:part_inter}
\end{figure}

\begin{figure*}
\centering
\begin{overpic}[width=1\textwidth,tics=10, trim=0mm 0 0mm 0,clip]{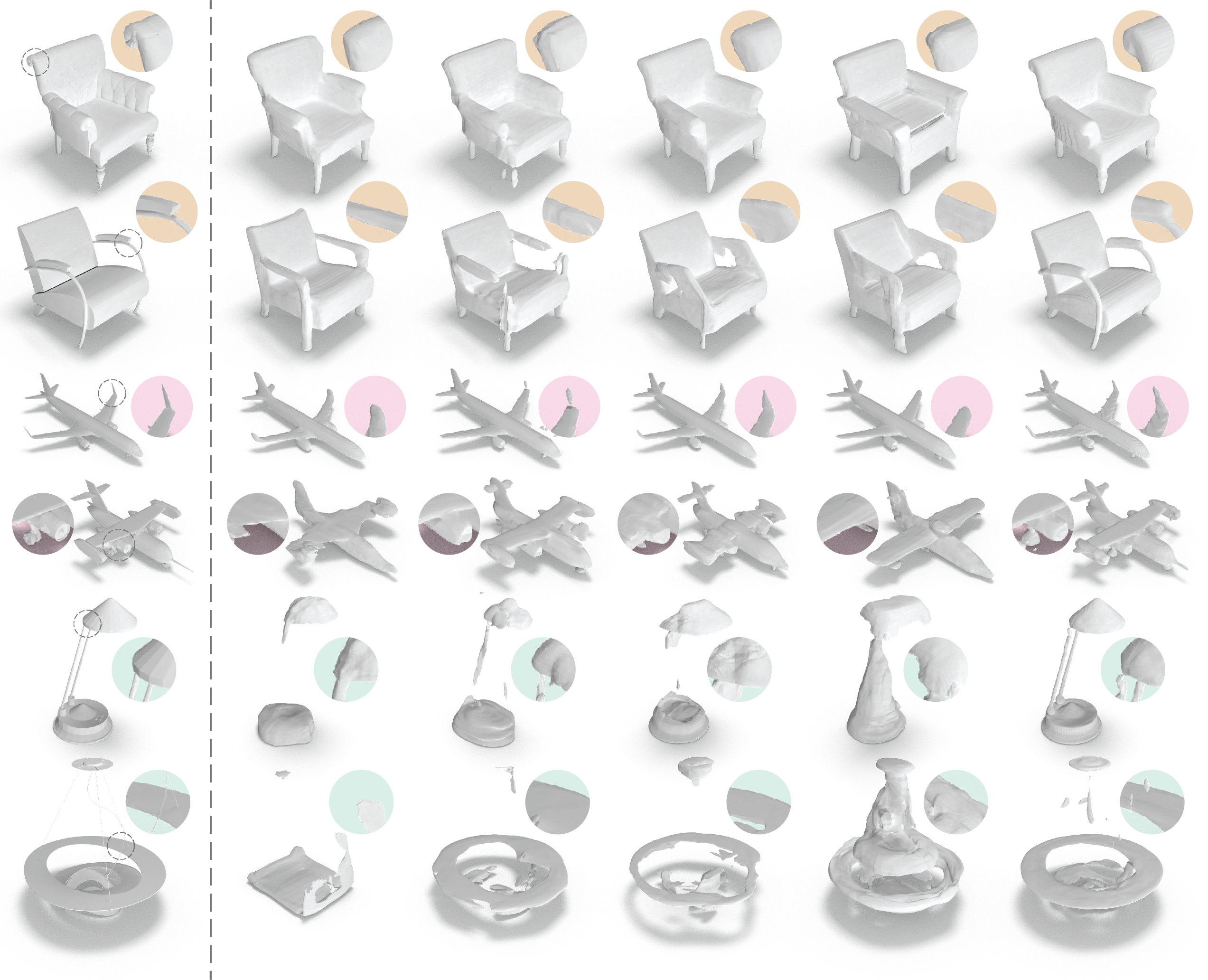}
    \put(3,0){Ground Truth}
    \put(22,0){IM-NET}
    \put(39.5,0){LDIF}
    \put(54.5,0){DeepSDF}
    \put(71,0){DualSDF}
    \put(86,0){\ourmethod{}}
    \end{overpic}

\caption{Shape inversion comparison. Uncurated results of the first two test shapes from the Chairs, Airplanes and Lamps categories of ShapeNet~\shortcite{chang2015shapenet}.}

\label{fig:inv_compare}
    
\end{figure*}
\begin{figure}
\centering
\begin{overpic}[width=1\columnwidth,tics=10, trim=0mm 0 0mm 0,clip]{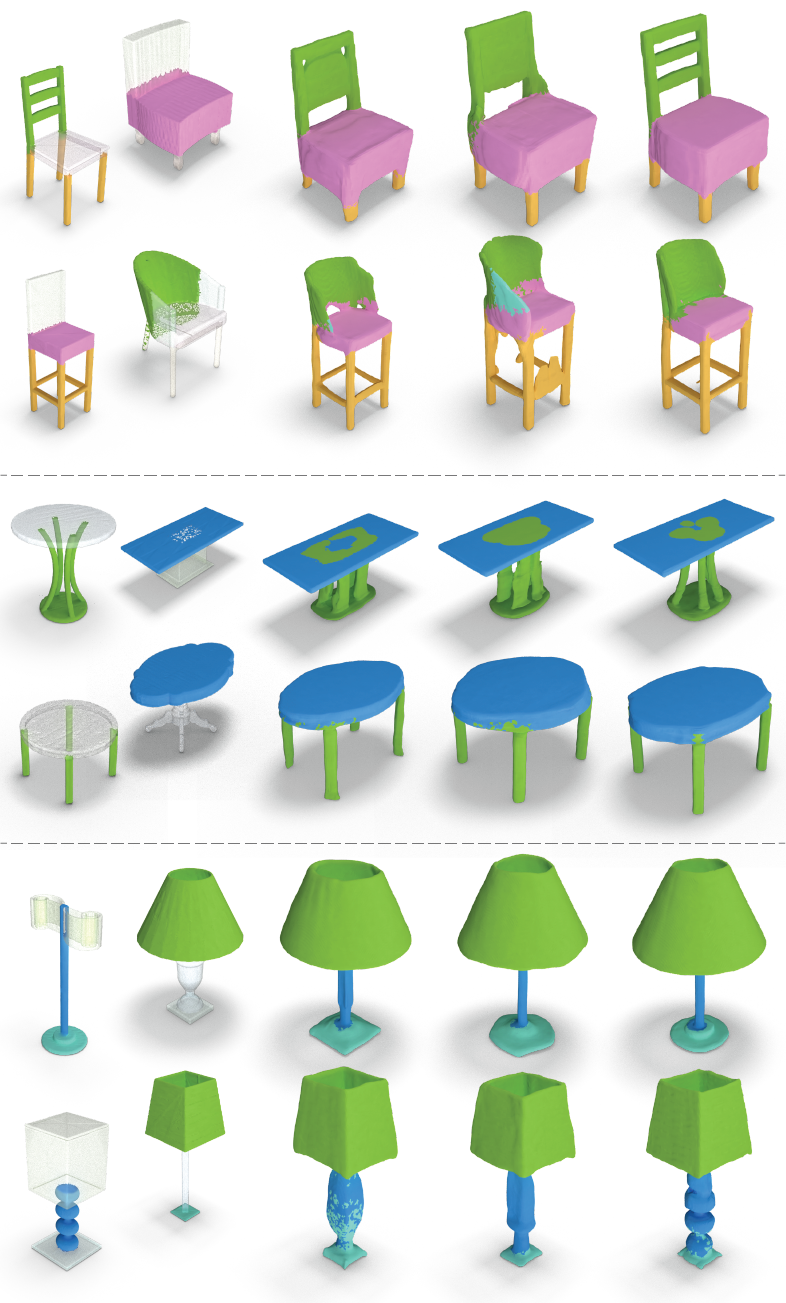}
    \put(6.5,-.5){Input parts}
    \put(22.5,-.5){no-dis}
    \put(37,-.5){no-enc}
    \put(52.5,-.5){full}
    \end{overpic}

\caption{Mixing ablation study. Each network configuration receives the selected parts (left) and outputs the mixed implicit shape (right).}

\label{fig:abl_mix}
    
\end{figure}
\section{Experiments}

\ourmethod{} can be leveraged for various applications that include shape inversion, generation, editing and mixing. These tasks are summarized in Table~\ref{tab:app}. In this section we demonstrate the advantages of \ourmethod{} for each application  and compare it to the other relevant learning based methods for implicit shapes synthesis.
In addition, we conduct an ablation study to evaluate the different components used in our framework.

Our experiments are conducted on the ShapeNet dataset \cite{chang2015shapenet}. We use the train-test split of DeepSDF \cite{park2019deepsdf}, where the number of shapes in the train set, per category, varies from $\sim 1k$ (lamps) to $\sim 5k$ (tables). Further details are included in Appendix~\ref{sec:imp_data}.

\subsection{Editing comparisons}
Concerning neural implicit shapes, only few methods provide editing controls. Moreover, some of the existing methods have set different editing objectives than us. In the following, we discuss the main difference between \ourmethod{} and two contemporary methods which are closest to us: COALESCE \cite{yin2020coalesce} and DualSDF \cite{hao2020dualsdf}. These works enable shape mixing and shape editing, respectively.

\textit{Shape mixing.} 
The objective of COALESCE is to put together a given set of shape parts and output a unified shape. Their pipeline consists of three main steps: (i) A part alignment network, which outputs a transformation for each input part such that the transformed parts are aligned together. (ii) A joint synthesis network that synthesizes new implicit joint parts, which connect the separate parts. (iii) "Poisson mesh stitching" is applied over the parts and the newly formed joints.
Each COALESCE network is trained over segmented parts from a single shape category.
Figure~\ref{fig:col_mix} shows mixing examples of our method compared to the stitching operation of COALESCE, using pre-trained networks on the chairs and airplanes categories.
Evidently, synthesizing joints as implicit functions and then stitching them \textit{automatically} to  existing meshes is prone to result in noisy artifacts at the joints regions. \ourmethod{} combines the latent parts to synthesize the entire implicit shape as a single unit, which results in generated shapes appearing more globally coherent.

\textit{Shape editing.}
 Given a single latent vector, DualSDF simultaneously synthesizes a high-resolution implicit shape and a corresponding coarse representation, made out of primitive shapes. Then, the user can control the implicit shape by manipulating the primitives of the coarse representation. Each editing step is followed by an optimization of a global latent, which yields a new implicit shape, satisfying constraints set by the coarse form.
 Figure ~\ref{fig:local_edit} illustrates the differences in the characteristics of our method and DualSDF \cite{hao2020dualsdf}.
The coarse representation of DualSDF does not hold the full geometric properties of the corresponding implicit shape. As a result, after each editing step, the geometry of the shape, or even its topology, might differ from its starting form. Edit operations are not guaranteed to preserve the identity of the original shape, but rather allow a \textit{guided} traversal over the latent space.
\ourmethod{} is resilient to such effects as it is trained over an embedding space of disentangled parts. Consequentially, the effect of local changes is bounded, e.g, each editing step affects only a specified portion of the latent representation, therefore, not risking a change of the shape identity.

\subsection{Shape inversion}
\label{sec:invert_eval}
We evaluate the shape inversion quality of our method on $3$ shape categories from the Shapenet dataset \cite{chang2015shapenet}: airplanes, chairs and lamps, using the train-test split of DeepSDF \cite{park2019deepsdf}. 
We compare our results to other generative and reconstruction methods that output implicit shapes: IM-NET\cite{chen2019learning}, LDIF \cite{genova2019learning}, DeepSDF \cite{park2019deepsdf} and DualSDF \cite{hao2020dualsdf}. 
We train each method using their official code and training settings.

The shape inversion process for auto-decoder based methods, DeepSDF and DualSDF, is obtained through an optimization process. They optimize a latent shape vector $z$, that minimizes the reconstruction loss: 
$$
\argmin_{\mathbf{z}} \mathcal{L}_{rec}\left(G(x, \mathbf{z}), y\right),
$$
where $G$ is the method's generative model, $x$ are the $3D$ coordinates and  $y$ are their corresponding labels, i.e., the signed distances of the SDF representation. ${L}_{rec}$ is determined by the specific loss settings of each method.

IM-NET and LDIF use encoder-decoder architectures and achieve shape inversion by employing the encoder. IM-NET's occupancy network is conditioned on a latent vector that is encoded from a 3D occupancy grid of the input shape.
LDIF encodes $24$ depth images from different views of the input shape and recovers an implicit shape.

\textit{Evaluation metric.}
Following prior works, the measured distances between the \textit{inverted} implicit shape and the ground truth meshes are chamfer distance (mean and median), earth mover's distance (mean and median) and mesh accuracy \cite{seitz2006comparison}.
The chamfer distance and earth mover's distance are measured between $30,000$ and $1024$ sampled points, respectively, on the generated shape and the ground truth mesh.
The mesh accuracy value $d$ is the minimal  distance such that $90\%$ of $30,000$ sampled points on the generated shape are within an Euclidean distance $d$ of the surface of the ground truth shape. 

The quantitative results  are summarized in Table~\ref{tab:inv} and qualitative results are shown in Figure~\ref{fig:inv_compare}.

\subsection{Shape generation evaluation}
\label{sec:shape_gen}
In addition to the editing capabilities of our method, we can use a pre-trained \ourmethod{} network for random, unconditional shape generation.
Similar to \cite{bojanowski2018optimizing}, we represent the latent distribution as a multivariate Gaussian distribution that best fits the latent space $\zA$ of our training data. Then, for shape generation, we feed-forward a sampled vector $\mathbf{\za}$ through our network to get its corresponding shape. 

Ideally, for fair evaluation, we would like to measure the quality of our generated shapes set $A$, with respect to the underlying shape distribution of the training data. However, since this distribution is unknown, we can only measure the quality of the generated shapes in $A$ with respect to some empirical distribution represented by an additional shapes set $B$. In our case, the set $B$ is composed of shapes from the training and test datasets.

In our evaluations, we randomly sampled 2048 points on each shape in $A$ and $B$. Then, we followed prior 3D generative works ~\cite{yang2019pointflow, gal_iccvw21} and used the metrics introduced by \cite{achlioptas2018learning}:

\textbf{The Jensen-Shannon Divergence (JSD)} measured between the voxel occupancy probability induced by all shapes in $A$ versus all shapes in $B$.

\textbf{Coverage (Cov)} measured by the percentage of shapes in $B$ that are covered by a shape in $A$. For this evaluation we assign for each shape in $A$, its closest shape in $B$.  Then, a shape in $B$ is considered covered if it is assigned by at least one shape in $A$.

\textbf{Minimum matching distance (MMD)} measured by the average distance of each shape in $B$ and its \textit{closest} shape in $A$.

 \textbf{1-nearest neighbor accuracy (1-NNA)} proposed by~\cite{LopezPazO17}. This measurement 
penalizes each shape $s$ either in $A$ or $B$ whose \textit{closest} shape lays in the same group as $s$.

We show results with both chamfer distance (CD) and earth mover's distance (EMD) as the distance measures for the COV, MMD and 1-NNA metrics. We randomly generate $1000$ shapes to compose the set $A$ for each method and shape category.
The set $B$ is composed of randomly selected $500$ shapes from the training set and $500$ shapes from the test set, repeated per category.

The quantitative results  are summarized in Table~\ref{tab:gen}, where we compare our method to 3 other generative methods for implicit shapes: IM-NET\cite{chen2019learning}, DeepSDF \cite{park2019deepsdf} and DualSDF \cite{hao2020dualsdf}. We repeated our evaluation over 3 shape categories from the Shapenet dataset \cite{chang2015shapenet}: airplanes, chairs and tables. Additional qualitative results of random generated samples are included in appendix \ref{app:genetaion}.

\subsection{Ablation studies}
\label{sec:ablation}
To validate our architecture and training settings, we compare our final model to two reduced variants of our method. The first variant, "\noenc", omits the middle Mixing network $\netB$.
For the second variant, "\nodis", the network architecture remains the same, but we omit the disentanglement loss $\mathcal{L}_{dis}$ (Eq.~\eqref{eq:loss_dis}) from the training objective.

The results of the ablation study for shape inversion are included in Table~\ref{tab:inv}. Notably, the two reduced \ourmethod{} variants achieve slightly worse inversion results. More importantly, part level control and the quality of editable shapes significantly degrades for these reduced variants. In the following we further discuss these phenomena.

\textit{Disentanglement ablation.}
One of the main objectives of our work is to enforce part-level disentanglement over our mid latent representation $\zB$, such that shape manipulations achieved through modifications of latent vector $\zb_i$, will result only in \emph{local} changes to the output shape. Effectively, $\zb_i$ should control the shape region that is most likely coming from the corresponding Gaussian $g_i$.

We evaluate \ourmethod{}'s ability to correspond the extrinsic GMM representation with disjoint, implicit shape parts the network outputs, by conducting two ablation tests.

For each training shape, representations $\zB$ are obtained by feeding the shape through the Decomposition network. We then employ the coarse segmentation annotations of PartNet \cite{Mo_2019_CVPR}. We sample $1e6$ coordinates within the volume of the shape, and assign each of them the segmentation label from PartNet.

We compare every part from PartNet, against the Gaussians obtained from $\zB$.
Each segmented point is attributed to the Gaussian that maximizes its expectation (Eq.~\eqref{eq:re-label}). Then, each Gaussian is mapped to a PartNet label, according to the segmented points attributed to it.

Our setup so far, automatically simulates a user's selection of semantic parts by manually marking their corresponding Gaussians. We partitioned the mid-latent part representations $\zB$, and their corresponding GMM to disjoint groups $\zB = \bigcupdot_{i=1}^{p} Z^{b_-}_i$, where the Gaussians that correspond to each subset $Z^{b_-}_i$ represent a distinct segmented part $p_i$ among the $p$ parts provided by PartNet. 

Using this partitioned latent space, we measure two intersection over union (IoU) scores with respect to the $1e6$ points already sampled inside the shape, and $1e6$ more points uniformly sampled within the unit cube.

First, we measure the latent representations to part correspondence (Cor.) of each part \emph{separately}. For each part $p_i$, we mask out the representations in $\zB$ of the Gaussians not associated with $Z^{b_-}_i$. Then, we feed the $2e6$ sampled coordinates through the Occupancy network $\netC$, attended over masked part representations, processed by $\netB$. We specify a label of $y=1$ for each coordinate $\xcoord \in p_i$, and $y=0$ otherwise. The IoU is measured by comparing these labels with predictions obtained from the occupancy indicator $\netC\left(\xcoord | \netB \left( Z^{b_-}_i \right) \right)$. Intuitively, a higher IoU score indicates that the latent vectors $Z^{b_-}$ are indeed responsible for the generation of the specific parts associated with their Gaussians.

In the second test we conduct, we measure the coverage IoU (Cov.) between the union of implicit shapes obtained from the distinct parts and the full implicit shape. Here, a higher score means that the union of the separate implicit functions faithfully represents the complete shape, i.e., that:

$$
\bigcup_{i=1}^{p} \netC\left(\xcoord | \netB \left(Z^{b_-}_i\right)\right) \approx \netC\left(\xcoord | \netB \left(\zB \right)\right).
$$

We train and evaluate the three network configurations over three ShapeNet categories with part segmentation annotations: tables, chairs and lamps. The results are summarized in Table~\ref{tab:abl_cor}.

Compared to the full \ourmethod{} architecture, the "\ourmethod{} no-dis" variant, trained without the explicit disentanglement loss, yields poor correspondence between the latent part representations and the output shape.

"\ourmethod{} no-enc" variant, which omits the Mixing network, achieves slightly better IoU scores for local part correspondences (Cor.) but performs worse in terms of global shape coverage (Cov.). We attribute these differences to interactions that occur within the Mixing network, which augment the contextual representation of each part vector $\zc$ with global information. Introducing global information requires additional capacity from the part embeddings, which may come at the expanse of local part information. Nevertheless, in practice, we aim to maintain a balance between global coherency and local parts separation. We therefore find that the inclusion of the Mixing network is crucial for synthesizing distinct parts to a globally coherent shape (Figure~{\ref{fig:abl_mix}}). In particular, the Mixing network is an important backbone for the "mixing" edit-operations, e.g: synthesizing together parts from different shapes.

\textit{Mixing ablation.}
We conduct an additional ablation test, where we start with the same segmented part partition as before. Each latent code $\zb$ and its projected Gaussian are mapped to some part $p_i$ from PartNet, but for each shape, we replace one of the part codes $\zb$ with a code from another shape. The replaced code comes from the same part category, i.e: we may replace the latent code representing a leg of a chair with code representing a leg of another chair.

A qualitative comparison for this experiment is shown in Figure~\ref{fig:abl_mix}. We observe that the settings that remove the Mixing network (\ourmethod{} no-enc) or train without $\mathcal{L}_{dis}$ (\ourmethod{} no-dis) are prone to noisy artifacts. At the same time, our full settings preserve the distinct parts better, while generating well-figured mixed shapes.

Since we do not have a ground truth mixed shape to compare to, we are compelled to conduct an indirect quantitative comparison instead. Our ablation aims to give an upper bound for the quality of mixed shapes, by comparing specific attributes of the "mixed" shapes with respect to their input parts. 

The first approximation bound we use, measures the difference between the surface area of the mixed shape to the surface area of the input shape. We report the percentage of that difference with respect to the surface area of the input parts. We refer to this measurement as \emph{area evaluation}. Since this evaluation only gives a rough indication for the quality of mixed shapes, we suggest an additional criterion.

The second bound we employ, measures the segmentation quality of the generated shape with respect to the input parts.
For this evaluation, we trained a segmentation network \cite{qi2017pointnetplusplus} for each shape category, and used it to estimate the surface area of each segmentation class, per shape. Then, we divide those areas by the total area of the entire shape, to obtain a distribution over the segmentation-classes. We report the Jensen-Shannon divergence of this distribution, with respect to the ground truth segmentation distribution of the input parts. We refer to this measure as \emph{segmentation evaluation}.

Qualitative segmentation examples are shown in Figure~\ref{fig:abl_mix} and quantitative results are summarized in Table~\ref{tab:abl_mix} (Seg. and area evaluation).

\subsection{Part level interpolation}

We now demonstrate the properties of the obtained manifold that contains latent part embeddings $\zB$, and evaluate the quality of continuous interpolations between them. Our goal is to demonstrate that this manifold is smooth in the geometric sense.

Our evaluation does not concern global interpolation, where one can simply interpolate between different shape embeddings $\mathbf{\za}$ (although our method can be used also for this purpose as well), but instead we focus on interpolations performed between sets of shape parts. Such  interpolations are non-trivial, as given two part embeddings sets $\zB_1$ and $\zB_2$, we do not posses correspondences between matching part. Moreover, $\zB_1$ and $\zB_2$ might consist of different number of part embeddings. Therefore, we suggest an interpolation scheme throughout the attention weights computed by the cross attention layers of $\netC{}$.

First, we compute all contextual part embeddings for two given shapes: $\zC_1 = \netB(\zB_1)$ and $\zC_2 = \netB(\zB_2)$.
Then, we replace each multi head attention value with an interpolated attention weight.

Recall, the Occupancy network uses the multi head attention formulation: $\textrm{Attention}(\textbf{q}_d, K_d, V_d)$.
This is the Transformer decoder depicted in Eq.~\eqref{eq:cross_terms}, where $\textbf{q}_d$ is the query vector of $\hat{x_t}$, and $K_d$, $V_d$ are the key and value matrices for $\zC$. For brevity, we omit the decoder notation $d$ in the following description.

Let $K_1$, $K_2$ and $V_1$, $V_2$ represent the keys and values of $\zC_1$ and $\zC_2$, respectively (Eq.~\eqref{eq:cross_terms}). For a linear interpolation weight $\alpha \in [0, 1]$, the interpolated attention is given by:
\begin{equation*}
\begin{split}
\textrm{I-Attention}(\alpha, \textbf{q}, K_1, V_1, K_2, V_2) = (1 - \alpha) & \textrm{Attention}(\textbf{q}, K_1, V_1) + \\
 \alpha & \textrm{Attention}(\textbf{q}, K_2, V_2).
\end{split}
\end{equation*}
From here, we continue to output the occupancy indication by attending coordinates $\xcoord$ over the interpolated attention value.

Figure~\ref{fig:part_inter} illustrates some qualitative results of part level interpolation. During rendering, we highlight the interpolated parts by examining the attention weights of coordinates on the iso-surface, or vertices of a reconstructed mesh. We assign each surface coordinate $\xcoord$ to the part that \textit{receives} the most attention from $\xcoord$: $$
\argmax_{j} \sum_{t=0}^T \sum_{i=0}^h \textrm{softmax}\left(\dfrac{\mathbf{q_{ti}}K_{ti}^T}{\sqrt{d_k}} \right)_j,
$$
where aggregation is done over the transformer layers and heads. Finally, we color the vertices that are assigned to the symmetric difference indices $\mathbb{I}\left( \zC_1 \bigtriangleup \zC_2\right)$, that is, coordinates that attend to interpolated codes.

\begin{figure}
\centering
\includegraphics[width=\columnwidth]{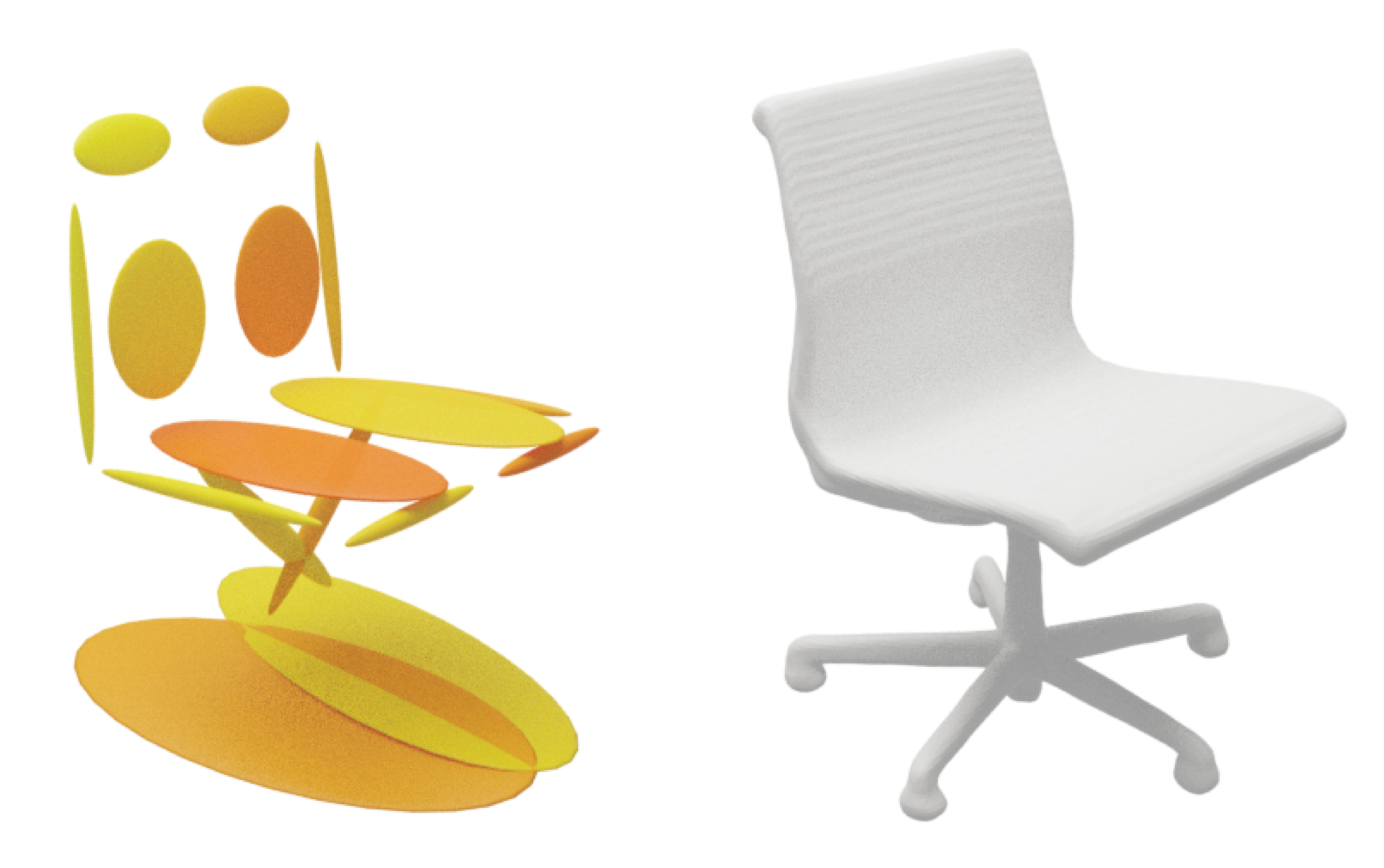}
\caption{Limitations. Since \ourmethod{} is trained without semantic parts supervision, the learned partitions may under-cluster semantic parts (the chair's back) , or limit the resolution of editing by over-clustering parts (the chair's leg).}

\label{fig:limitation}
\end{figure}

\section{Conclusions, limitations and future work}

We presented a generative neural model for implicit 3D shapes, featuring local editing capabilities. Our model is part-aware, requires no part-supervision, and leverages the Transformer architecture to form globally coherent shapes. The network was designed to allow editing at an interactive rate, where, as demonstrated, the user can interact with the generated model using a simple interface.

Our method relies on dual-level disentanglements. First, our model creates disentangled latent codes for disjoint generated shape parts. This allows supporting selection and mixing between different shape parts. Second, each part embedding is a factorized representation of intrinsic and extrinsic information, which are used to conduct local deformations over the shape. 

In our work, we assume to have 3D training data that consists of shapes with similar structure, such as chairs and airplanes. Such structured data enables the unsupervised learning of consistent partitions of compact parts. However, in some shape categories, this assumption holds weakly. For example, the lamps dataset consists of many unique lamps (Figure~\ref{fig:gmms}).
Moreover, since we do not employ part-level supervision, our model is agnostic to the semantics of part shapes. Our GMM partition may over-cluster together parts into the same Gaussian which may prevent the desired level of control. For example, see Figure \ref{fig:limitation}, where the leg of the chair is represented by only two Gaussians, even though it consists of many sub parts. Similarly, the model may under-cluster large parts and represent them with multiple latent codes, for example, the back of a chair can be represented by more than a single Gaussian.
We conjecture that these limitations can be addressed by introducing semi-supervised annotations, or through hierarchical part decompositions \cite{eckart2018hgmr}.

In this work, we explored how learning-based techniques can further assist future workflows of 3D modeling. In a supervised or semi-supervised settings, where the part decomposition is guided by instance-level labels, the performance can be further improved, which may also lead to co-segmentation as a byproduct.
In the future, we would like to add more intuitive 3D editing tools, and other types of interactive interfaces. For example, guiding the 3D modeling process by 2D sketches or through textual descriptions.

\bibliographystyle{ACM-Reference-Format}
\bibliography{ref}

\appendix

\section{Implementation Details}
\label{sec:imp} 

Our network architecture and training are implemented using Pytorch \cite{NEURIPS2019_9015}.
Our user interface is implemented using the  Visualization Toolkit (VTK).\footnote{ \url{http://www.vtk.org}}
\subsection{Data Preparation.}
\label{sec:imp_data}

We use the ShapeNet dataset \cite{chang2015shapenet} for training and testing our network to produce the results in this paper. Specifically, we use the train-test split of DeepSDF \cite{park2019deepsdf}. The number of training shapes in each category varies from $\sim$$1k$ (lamps) to $\sim$$5k$ (tables).

For each shape we sample points and occupancy labels by the following steps:
i) We convert the shape to a watertight mesh using the implementation of \citet{huang2018robust}.
ii) We scale the shape to a unit sphere. 
iii) For the GMM loss ($\xin$ in Eq.~\eqref{eq:loss_gmm}), we sample $500,000$ points uniformly inside each shape.
iv) For the occupancy loss  ($\xbatch$ in Eq.~\eqref{eq:loss_occ}), we randomly sample $500,000$ points inside the bounding cube $[-1, 1]^3$. Additional $500,000$ points are sampled on the surface of the mesh, and are perturbated with Gaussian noise of $\sigma = 0.01$. $500,000$ more surface points are perturbated with  $\sigma = 0.02$. In total, we sample  $1,500,000$ points for each shape.

For the occupancy labels we use the \emph{libigl} implementation \cite{libigl} of fast winding number \cite{Barill:FW:2018}.

Notice that in all the other methods that we compare to \cite{park2019deepsdf, chen2019learning, hao2020dualsdf, genova2019learning}, we use the same train/test shapes and use their official implementation\footnote{IM-NET \url{https://github.com/czq142857/IM-NET-pytorch} \\ DeepSDF \url{https://github.com/facebookresearch/DeepSDF} \\
DualSDF \url{https://github.com/zekunhao1995/DualSDF} \\
LDIF \url{https://github.com/google/ldif}} for both data preparation and training.

\subsection{Network Architecture}
\label{sec:imp_net}
Throughout all experiments, our networks were trained using the settings below.
 
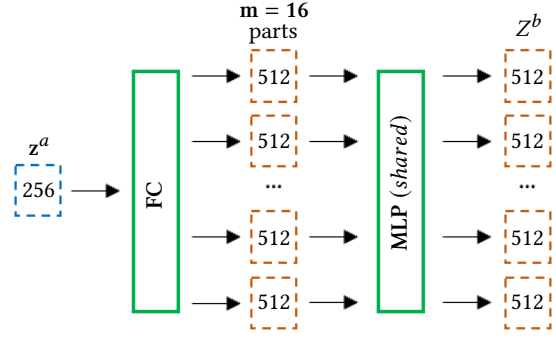
\begin{figure}
\centering
\begin{overpic}[width=1\columnwidth,tics=10, trim=0mm 0 0mm 0,clip]{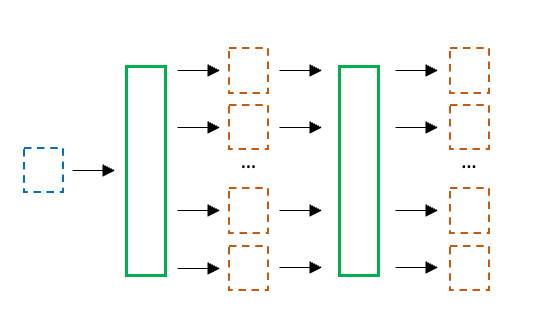}
    \put(6.2,35){$\za$}
    \put(5.2,28.5){$256$}
    
    \put(24.5,26.5){\rotatebox{90}{$\mathbf{FC}$}}
    
    \put(39,56.5){$\mathbf{m=16}$}
    \put(40.5,53.2){parts}
    \put(41.8,46){$512$}
    \put(41.8,36){$512$}
    \put(41.8,21){$512$}
    \put(41.8,11){$512$}
    
    \put(62.4,20.0){\rotatebox{90}{$\mathbf{MLP} \ (shared)$}}
    
    \put(82,53.5){$\zB$}
    \put(81.2,46){$512$}
    \put(81.2,36){$512$}
    \put(81.2,21){$512$}
    \put(81.2,11){$512$}
    
    \end{overpic}
\caption{Architecture of Decomposition network $\netA$. Shape embedding $\za$ is projected through a fully connected layer to $m$ intermediate part embeddings of dimension 512 each. Then each embedding is forwarded through a shared MLP with one hidden layer to produce output part embedding $\zb \in \zB$ of 512 dimensions.}
\label{fig:decomp_arch}
\vspace{-2mm}
\end{figure}

\textbf{Decomposition Network}
As illustrated in Figure~\ref{fig:decomp_arch},
the dimension of the shape codes $\zA$ is 256. Shape codes are initialized randomly from $\mathcal{N}\left(0, 256^{-2}\right)$. 
The first fully connected layer of $\netA$ splits $\za$ into  $m=16$ part vectors of dimension $512$ each. 
We then apply a MLP with one hidden layer of dimension $1024$, and output representations $\zB \in \reals^{m \times 512}$.

\textbf{Implicit Shape Composer.}
The Mixing network $\netB$ and Occupancy network $\netC$ are utilized through the full Transformer architecture \cite{vaswani2017attention}. We removed the self attention layers from the decoder, and used learned positional encoding for the decoder as well. These changes are described in details in Section~$\ref{sec:composition}$.

We use $\dm=512$ as the dimension of the Mixing network (Transformer Encoder).
For the Occupancy network (Transformer Decoder), we set the dimension of $\dpe$ to $256$.
Both, transformer encoder and decoder, have $T=4$ multi-head attention layers with $h=8$ number of heads.

To output the occupancy indicator $\hat{y}$, we process $\xcoordhat_{T} \in \reals^{\dpe}$ through another MLP with two hidden layers of size $512$.

\subsection{Training}
\label{sec:imp_opt}
Each network was trained to minimize the loss term in Eq.~\eqref{eq:loss_all} for $2000$ epochs with batch size of $18$ shapes. Each shape in the batch is represented by $2000$ uniformly sampled points inside $\xin$, and $6000$ occupancy points of $\xbatch$, where both are randomly selected from the pre-processed points, as described in \ref{sec:imp_data}.

We set the loss weight in Eq.~\eqref{eq:loss_all} to $\gamma = 10^{-4}$. We use the Adam optimizer \cite{kingma2014adam} with a learning rate of $10^{-4}$ and the default settings ($\beta_1 = 0.9, \beta_2 = 0.999, \epsilon= 10^{-8}$). We use $2000$ warm-up iterations and apply exponential learning decay of $0.9$  in intervals of $500$ epochs.

The affine transformations that are applied on $\xbatch$ and $\xbatch^\textrm{-}$ (see Section~\ref{sec:att_disentanglement}) are randomly selected from $100,000$ pre-computed transformations. Each is composed from a random translation $t \in [-0.3, 0.3]^3$, random uniform scale $s \in [0.7, 1.3]$ and a random uniform rotation from $SO(3)$.

Each model was trained on a single RTX A6000 GPU for 1-5 days, depending on the size of the training data. We run our user interface on a laptop equipped with RTX 3700 GPU.

\subsection{Shape Inversion Optimization}
\label{sec:imp_shape_inversion}

For shape inversion, we perform a 2-step optimization, as described in Section~\ref{sec:invert}. Both the first and second steps of the optimization are run for a fixed amount of 250 iterations. 

On a machine equipped with a RTX 3700 GPU, the processing time of 800 shapes takes ~2 hours, where shapes are processed in batches of 20 in parallel.

\opc{TODO:
1. Training time - V, optimization time for inversion, unconditional generation time

2. Stopping criteria for optimizations: how many epochs?
}

\section{Random Generation Results}
\label{app:genetaion}

Below are qualitative comparisons for the random generation evaluation in Section~\ref{sec:shape_gen}.
For each method and shape category we show the first $36$ generated shapes.
\begin{figure*}
\centering

\begin{overpic}[width=1\textwidth,tics=10, trim=0mm 0 0mm 0,clip]{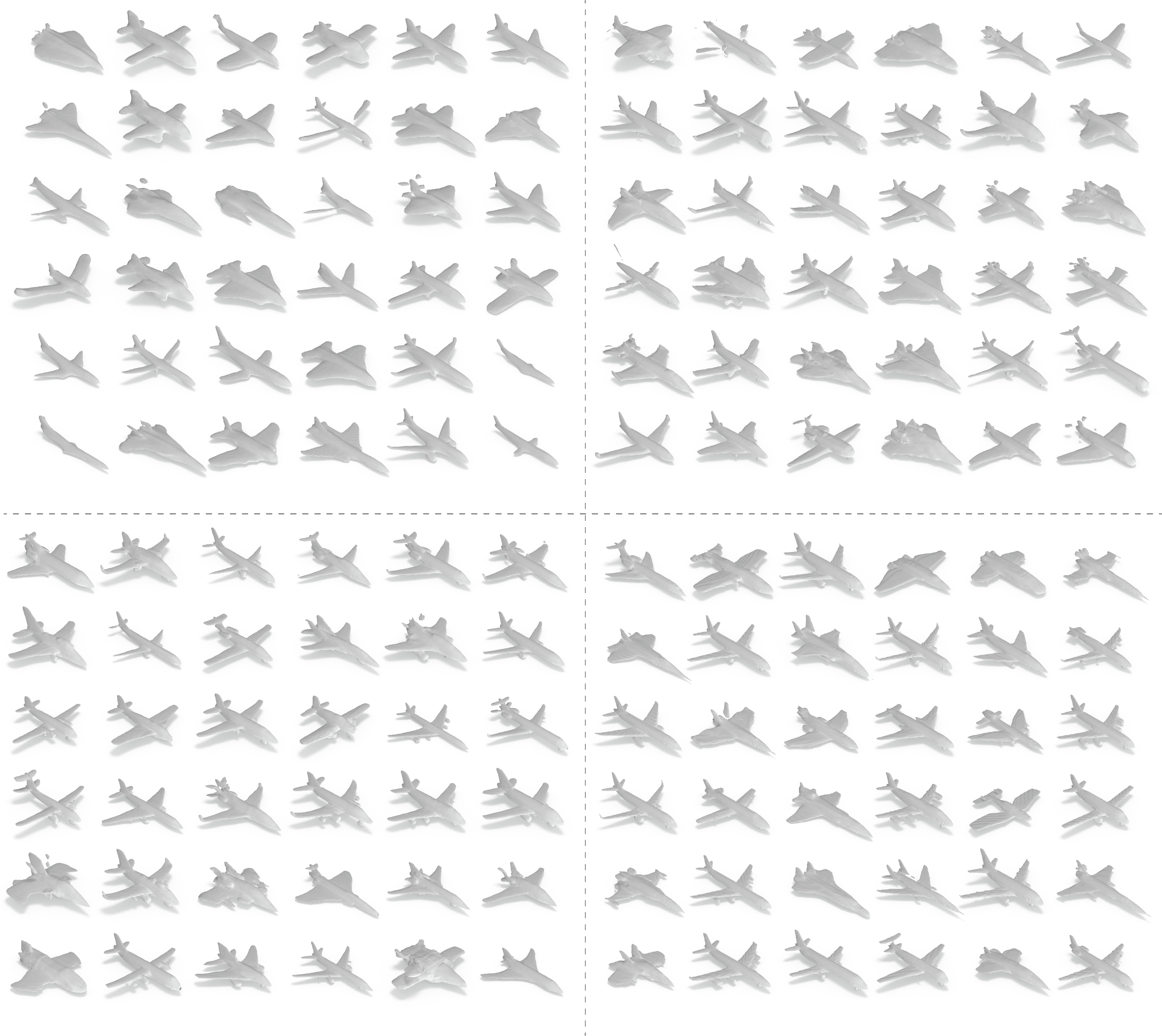}
    \put(22,46){IM-NET}
    \put(72,46){DeepSDF}
    \put(22,0){DualSDF}
    \put(71,0){\ourmethod{}}
    \end{overpic}

\caption{Shape generation results. First 36 sampled airplanes used for the comparison in Table~\ref{tab:gen}.}

\label{fig:gen_airplanes}
    
\end{figure*}

\begin{figure*}
\centering

\begin{overpic}[width=1\textwidth,tics=10, trim=0mm 0 0mm 0,clip]{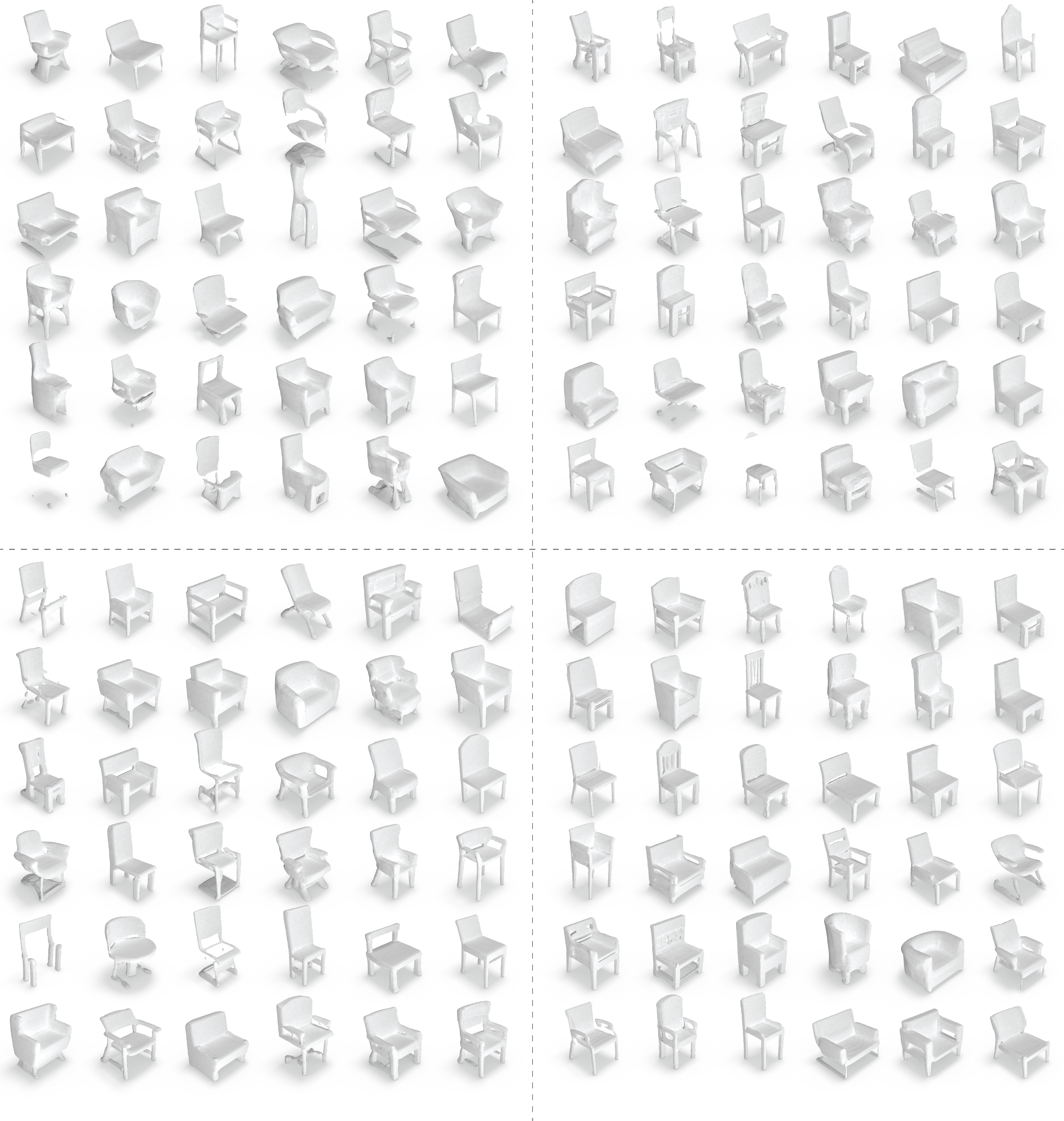}
    \put(20,52){IM-NET}
    \put(70,52){DeepSDF}
    \put(19,0){DualSDF}
    \put(67,0){\ourmethod{}}
    \end{overpic}

\caption{Shape generation results. First 36 sampled chairs used for the comparison in Table~\ref{tab:gen}.}

\label{fig:gen_chairs}
    
\end{figure*}

\begin{figure*}
\centering

\begin{overpic}[width=1\textwidth,tics=10, trim=0mm 0 0mm 0,clip]{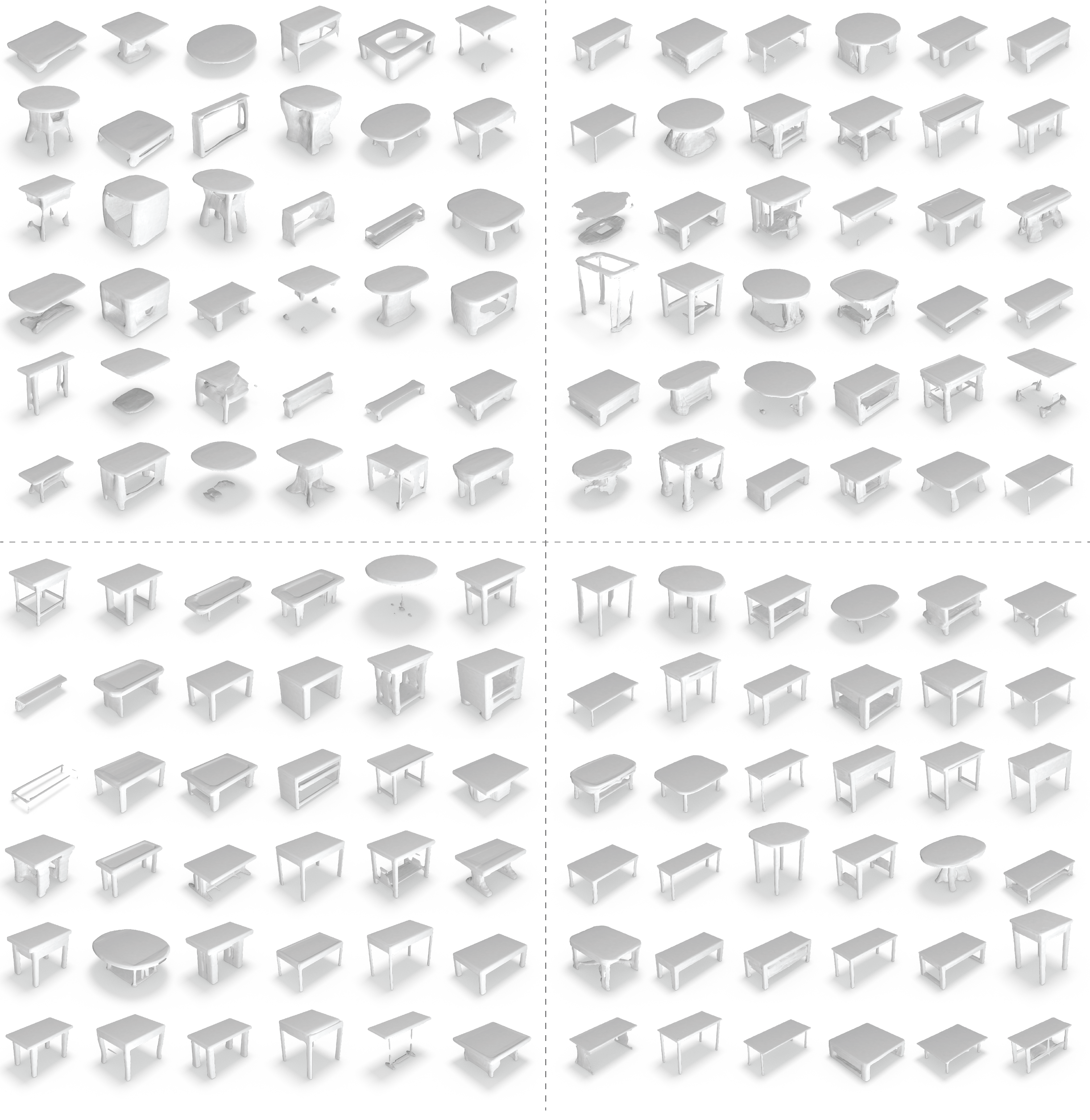}
    \put(20,52.5){IM-NET}
    \put(69,52.5){DeepSDF}
    \put(20,0){DualSDF}
    \put(68,0){\ourmethod{}}
    \end{overpic}

\caption{Shape generation results. First 36 sampled tables used for the comparison in Table~\ref{tab:gen}.}

\label{fig:gen_tables}
    
\end{figure*}
\end{document}